\documentclass[10pt,twocolumn,twoside,journal]{IEEEtran}
\usepackage{subfigure}
\usepackage{setspace}
\usepackage{amsmath}
\usepackage{amssymb}
\usepackage{amsfonts}
\usepackage{amscd}
\usepackage{mathrsfs}
\usepackage[final]{graphicx}
\usepackage{graphicx}
\usepackage{psfrag}
\usepackage{url}
\usepackage{textcomp}
\usepackage{multirow}

\newcommand{\Define}{\stackrel{\triangle}{=}}

\newtheorem{thm}{\bf Theorem}

\begin{document}

\markboth{Accepted in IEEE JOURNAL OF SELECTED TOPICS IN SIGNAL PROCESSING, 
Vol. XX, 
No. XX, XXXX}{Saif K. Mohammed \MakeLowercase{\textit{et al.}}: High-Rate 
Space-Time Coded Large-MIMO Systems: Low-Complexity Detection and Channel 
Estimation}

\title{High-Rate Space-Time Coded Large-MIMO Systems: Low-Complexity
Detection and \\ Channel Estimation}
\author{Saif K. Mohammed, Ahmed Zaki, A. Chockalingam, \IEEEmembership{Senior 
Member, IEEE}, and B. Sundar Rajan, \IEEEmembership{Senior Member, IEEE}
\thanks{This work in part was presented in IEEE ISIT'2008, Toronto, Canada,
July 2008, in IEEE GLOBECOM'2008, New Orleans, USA, December 2008, and in
IEEE ICC'2009, Dresden, Germany, June 2009. 
The authors are with the Department of Electrical Communication 
Engineering, Indian Institute of Science, Bangalore-560012, India.
E-mail: saifind2007@yahoo.com, zakismail@gmail.com, 
\{achockal,bsrajan\}@ece.iisc.ernet.in}
\thanks{Manuscript received August 2008; revised March 23, 2009.}}

\markboth{Accepted in IEEE JOURNAL OF SELECTED TOPICS IN SIGNAL PROCESSING: 
Spl. Iss. on Managing Complexity in Multiuser MIMO Systems}{Saif
K. Mohammed \MakeLowercase{\textit{et al.}}: High-Rate Space-Time Coded 
Large-MIMO Systems: Low-Complexity Detection and Channel Estimation}

\maketitle

\begin{abstract}
In this paper, we present a low-complexity algorithm for detection in
high-rate, non-orthogonal space-time block coded (STBC) large-MIMO systems
that achieve high spectral efficiencies of the order of tens of bps/Hz. We
also present a training-based iterative detection/channel estimation scheme
for such large STBC MIMO systems. Our simulation results show that excellent
bit error rate and nearness-to-capacity performance are achieved by the
proposed multistage {\em likelihood ascent search} ($M$-LAS) detector in 
conjunction with the proposed iterative detection/channel estimation scheme 
at low complexities. The fact that we could show such good results for large 
STBCs like $16\times 16$ and $32\times 32$ STBCs from Cyclic Division Algebras
(CDA) operating at spectral efficiencies in excess of 20 bps/Hz (even after
accounting for the overheads meant for pilot based training for channel
estimation and turbo coding) establishes the effectiveness of the proposed
detector and channel estimator. We decode perfect codes of large dimensions
using the proposed detector. With the feasibility of such a low-complexity
detection/channel estimation scheme, large-MIMO systems with tens of antennas
operating at several tens of bps/Hz spectral efficiencies can become practical, 
enabling interesting high data rate wireless applications.
\end{abstract}


\section{Introduction}
\label{sec1}
Current wireless standards (e.g., IEEE 802.11n and 802.16e) have adopted
MIMO techniques \cite{tela99}-\cite{jafarkhani} to achieve the benefits
of transmit diversity (using space-time coding) and high data rates (using
spatial multiplexing). They, however, harness only a limited potential of
MIMO benefits since they use only a small number of transmit
antennas (e.g., 2 to 4
antennas). Significant benefits can be realized if large number of antennas
are used; e.g., large-MIMO systems with tens of antennas in communication
terminals can enable multi-giga bit rate transmissions at high spectral
efficiencies of the order of {\em several tens of bps/Hz}\footnote{Spectral
efficiencies achieved in current MIMO wireless standards are only about
10 bps/Hz or less.}. Key challenges in realizing such large-MIMO systems
include low-complexity detection and channel estimation, RF/IF technologies,
and placement of large number of antennas in communication
terminals\footnote{WiFi products in 2.5 GHz band which use 12 transmit 
antennas for beamforming purposes are becoming commercially available 
\cite{ruckus}. With such RF and antenna technologies for placing large 
number of antennas in medium/large aperture communication terminals (like
set-top boxes/laptops) getting increasingly matured, low-complexity
high-performance MIMO baseband receiver techniques (e.g., detection
and channel estimation) are crucial to enable practical implementations
of high spectral efficiency large-MIMO systems, which, in turn, can
enable high data rate applications like wireless IPTV/HDTV distribution.}.
Our focus in this paper is on low-complexity detection and channel
estimation for large-MIMO systems.

Spatial multiplexing (V-BLAST) with large number of transmit
antennas can offer
high spectral efficiencies, but it does not give transmit diversity.
On the other hand, well known orthogonal space-time block codes (STBC)
have the advantages of full transmit diversity and low decoding complexity,
but they suffer from rate loss for increasing number of transmit antennas
\cite{jafarkhani},\cite{Alamouti},\cite{tarokh3}. However, {\em full-rate,
non-orthogonal STBCs from Cyclic Division Algebras (CDA)} \cite{bsr} are
attractive to achieve high spectral efficiencies in addition to achieving
full transmit diversity, using large number of transmit antennas. For
example, a $32\times 32$ STBC matrix from CDA has 1024 symbols (i.e., 32
complex symbols per channel use), and using this STBC along with 16-QAM
and rate-3/4 turbo code offers a spectral efficiency of 96 bps/Hz. While
maximum-likelihood (ML) decoding of orthogonal STBCs can be achieved in
linear complexity, ML or near-ML decoding of non-orthogonal STBCs with
large number of antennas at low complexities has been a challenge. Channel
estimation is also a key issue in large-MIMO systems. In this paper, we
address these two challenging problems; our proposed solutions can
potentially enable realization of large-MIMO systems in practice.

Sphere decoding and several of its low-complexity variants are known in the
literature \cite{sphere0}-\cite{spherex}. These detectors, however, are
prohibitively complex for large number of antennas. Recent approaches to
low-complexity multiuser/MIMO detection involve application of techniques
from belief propagation \cite{bp1}, Markov Chain Monte-Carlo methods
\cite{mcmc}, neural networks \cite{sun},\cite{jsac},\cite{icc08}, etc.
In particular, in \cite{jsac},\cite{icc08}, we presented a powerful
Hopfield neural network based low-complexity search algorithm for
detecting large-MIMO V-BLAST signals, and showed that it performs
quite close to (within 4.6 dB of) the theoretical capacity, at
high spectral efficiencies of the order of tens to hundreds of bps/Hz
using tens to hundreds of antennas, at an average per-symbol detection
complexity of just $O(N_tN_r)$, where $N_t$ and $N_r$ denote the number
of transmit and receive antennas, respectively.

In this paper, we present $i$) a low-complexity near-ML achieving
detector, and $ii$) an iterative detection/channel estimation scheme
for large non-orthogonal STBC MIMO systems having tens of transmit 
and receive antennas.
Our key contributions here can be summarized as follows:
\begin{enumerate}
\item   We generalize the 1-symbol update based {\em likelihood ascent 
	search} (LAS) algorithm we proposed in \cite{jsac},\cite{icc08}, 
	by employing a low-complexity multistage multi-symbol update based
        strategy; we refer to this new algorithm as multistage LAS ($M$-LAS) 
	algorithm. We show that the $M$-LAS algorithm outperforms the basic 
	LAS algorithm with some increase in complexity.

\item   We propose a method to generate soft outputs from the $M$-LAS
        output vector. Soft outputs generation was not considered in
        \cite{jsac},\cite{icc08}. The proposed soft outputs generation
        for the individual bits results in about 1 to 1.5 dB improvement
        in coded bit error rate (BER) compared to hard decision $M$-LAS 
	outputs.

\item   Assuming i.i.d. fading and perfect channel state information
        at the receiver (CSIR), our simulation results show that the
        proposed $M$-LAS algorithm is able to decode large non-orthogonal
        STBCs (e.g., $16\times 16$ and $32\times 32$ STBCs) and achieve
        near single-input single-output (SISO) AWGN uncoded BER performance 
	as well as near-capacity (within 4 dB from theoretical capacity) 
	coded BER performance.

\item   Using the proposed detector, we decode and report the simulated
        BER performance of `perfect codes' \cite{gold05}-\cite{cda} of
        large dimensions.

\item   Presenting a BER performance and complexity comparison of the
        proposed CDA STBC/$M$-LAS detection approach with other 
        large-MIMO/detector approaches (e.g., stacked Alamouti codes/QOSTBCs
        and associated interference canceling receivers reported in
        \cite{kazem}), we show that the proposed approach outperforms
        the other considered approaches, both in terms of performance as
        well as complexity.

\item   We present simulation results that quantify the loss in BER
        performance due to spatial correlation in large-MIMO systems, by
        considering a more realistic spatially correlated MIMO fading
        channel model proposed by Gesbert {\em et al} in \cite{mimo3}. We 
	show that this loss in performance can be alleviated by providing 
	more receive dimensions (i.e., more receive antennas than transmit
        antennas).

\item   Finally, we present a training-based iterative detection/channel
        estimation scheme for large STBC MIMO systems. We report
        BER and nearness-to-capacity results when the channel matrix is
        estimated using the proposed iterative scheme and compare these
        results with those obtained using perfect CSIR assumption.
\end{enumerate}
The rest of the paper is organized as follows. In Section \ref{sec2}, we
present the STBC MIMO system model considered. The proposed detection
algorithm is presented in Section \ref{sec3}. BER performance results with
perfect CSIR are presented in Section \ref{sec4}. This section includes
the results on the effect of spatial correlation, BER performance
of large perfect codes, and comparison of the proposed scheme with other
large-MIMO architecture/detector combinations. The proposed iterative
detection/channel estimation scheme and the corresponding performance
results are presented in Section \ref{sec5}. Conclusions are presented
in Section \ref{sec6}.

\section{System Model}
\label{sec2}
Consider a STBC MIMO system with multiple transmit and multiple receive
antennas. An $(n,p,k)$ STBC is represented by a matrix
{\small ${\bf X}_c \in {\mathbb C}^{n \times p}$}, where $n$ and $p$
denote the number of transmit antennas and number of time slots,
respectively, and $k$ denotes the number of complex data symbols sent
in one STBC matrix. The $(i,j)$th entry in ${\bf X}_c$ represents the
complex number transmitted from the $i$th transmit antenna in the $j$th
time slot. The rate of an STBC, $r$, is given by $r\ \Define \frac{k}{p}$.
Let $N_r$ and $N_t=n$ denote the number of receive and transmit antennas,
respectively. Let ${\bf H}_c \in {\mathbb C}^{N_r\times N_t}$ denote the
channel gain matrix, where the $(i,j)$th entry in ${\bf H}_c$ is the
complex channel gain from the $j$th transmit antenna to the $i$th receive
antenna. We assume that the channel gains remain constant over one STBC
matrix duration. Assuming rich scattering, we model the entries of
${\bf H}_c$ as i.i.d
$\mathcal C \mathcal N(0,1)$\footnote{${\cal CN}(0,\sigma^2)$ denotes a
circularly symmetric complex Gaussian distribution with mean zero and
variance $\sigma^2$.}.
The received space-time
signal matrix, ${\bf Y}_c \in {\mathbb C}^{N_r \times p}$, can be written as
\begin{eqnarray}
\label{SystemModel}
{\bf Y}_c & = & {\bf H}_c{\bf X}_c + {\bf N}_c,
\end{eqnarray}
where ${\bf N}_c \in {\mathbb C}^{N_r \times p}$ is the noise matrix at
the receiver and its entries are modeled as i.i.d
$\mathcal C \mathcal N\big(0,\sigma^2=\frac{N_tE_s}{\gamma}\big)$,
where $E_s$ is the average energy of the transmitted symbols, and
$\gamma$ is the average received SNR per receive antenna \cite{jafarkhani},
and the $(i,j)$th entry in ${\bf Y}_c$ is the received signal at the $i$th
receive antenna in the $j$th time slot. In a linear dispersion (LD)
STBC, ${\bf X}_c$ can be decomposed into a linear
combination of weight matrices corresponding to each data symbol and its
conjugate as \cite{jafarkhani}
\begin{eqnarray}
\label{SystemModelStbcLp}
{\bf X}_c & = & \sum_{i = 1}^{k} x_c^{(i)} {\bf A}_c^{(i)} + (x_c^{(i)})^* {\bf E}_c^{(i)},
\end{eqnarray}
where $x_c^{(i)}$ is the $i$th complex data symbol, and
${\bf A}_c^{(i)}, {\bf E}_c^{(i)} \in {\mathbb C}^{N_t \times p}$ are
its corresponding weight matrices. The detection algorithm we propose
in this paper can decode general LD STBCs of the form in
(\ref{SystemModelStbcLp}). For the purpose of simplicity in exposition,
here we consider a subclass of LD STBCs, where ${\bf X}_c$ can be written
in the form
\begin{eqnarray}
\label{SystemModelStbcLpx}
{\bf X}_c & = & \sum_{i = 1}^{k} x_c^{(i)} {\bf A}_c^{(i)}.
\end{eqnarray}
From (\ref{SystemModel}) and (\ref{SystemModelStbcLpx}), applying the
$vec\,(.)$ operation\footnote{For a $p\times q$ matrix ${\bf M}
= [{\bf m}_1 {\bf m}_2 \cdots {\bf m}_q]$, where ${\bf m}_i$ is
the $i$th column of ${\bf M}$, $vec({\bf M})$ is a $pq\times 1$
vector defined as
$vec({\bf M})=[{\bf m}_1^T {\bf m}_2^T \cdots {\bf m}_q^T]^T$, where
$[.]^T$ denotes the transpose operation.}
 we have
\begin{eqnarray}
\label{SystemModelvec}
vec\,({\bf Y}_c) & = & \sum_{i=1}^{k} x_c^{(i)} vec\,({\bf H}_c {\bf A}_c^{(i)}) + vec\,({\bf N}_c).
\end{eqnarray}
If {\small ${\bf U}$,${\bf V}$,${\bf W}$,${\bf D}$} are matrices such that
{\small ${\bf D}={\bf U}{\bf W}{\bf V}$}, then it is true that
{\small $vec\,({\bf D})=( {\bf V}^T \otimes {\bf U}) \, vec\,({\bf W})$},
where $\otimes$ denotes tensor product of matrices
\cite{matrix}. Using this, we can
write (\ref{SystemModelvec}) as
\begin{eqnarray}
\label{SystemModelvec1}
\hspace{-3mm}
vec\,({\bf Y}_c) & = & \sum_{i=1}^{k} x_c^{(i)} ({\bf I} \otimes {\bf H}_c)\, vec\,({\bf A}_c^{(i)}) + vec\,({\bf N}_c),
\end{eqnarray}
where ${\bf I}$ is the $p \times p$ identity matrix.
Further, define
${\bf y}_c \Define vec\,({\bf Y}_c)$,
$\widehat{{\bf H}}_c \Define ({\bf I} \otimes {\bf H}_c)$,
${\bf a}_c^{(i)} \Define vec\,({\bf A}_c^{(i)})$, and
${\bf n}_c \Define vec\,({\bf N}_c)$.
From these definitions, it is clear that
${\bf y}_c \in {\mathbb C}^{N_rp \times 1}$,
$\widehat{{\bf H}}_c \in {\mathbb C}^{N_rp  \times N_tp}$,
${\bf a}_c^{(i)} \in {\mathbb C}^{N_tp \times 1}$, and
${\bf n}_c \in {\mathbb C}^{N_rp \times 1}$.
Let us also define a matrix
$\widetilde{{\bf H}}_c \in {\mathbb C}^{N_rp \times k}$, whose
$i$th column is
$\widehat{{\bf H}}_c \, {\bf a}_c^{(i)}$, $i=1,\cdots,k$.
Let ${\bf x}_c \in {\mathbb C}^{k \times 1}$,
whose $i$th entry is the data symbol $x_c^{(i)}$.
With these definitions, we can write (\ref{SystemModelvec1}) as
\begin{eqnarray}
\label{SystemModelvec2}
{\bf y}_c & = & \sum_{i=1}^{k} x_c^{(i)}\, (\widehat{{\bf H}}_c\, {\bf a}_c^{(i)}) + {\bf n}_c \,\,\, = \,\,\, \widetilde{{\bf H}}_c {\bf x}_c + {\bf n}_c .
\end{eqnarray}
Each element of ${\bf x}_c$ is an ${\mathcal M}$-PAM or
${\mathcal M}$-QAM symbol. ${\mathcal M}$-PAM
symbols take discrete values from {\small $\{A_m, m=1,\cdots,{\mathcal M}\}$},
where $A_m=(2m-1-{\mathcal M})$, and ${\mathcal M}$-QAM is nothing but two
PAMs in quadrature.
Let ${\bf y}_c$, $\widetilde{{\bf H}}_c$, ${\bf x}_c$, and ${\bf n}_c$
be decomposed into real and imaginary parts as
\begin{eqnarray}
\label{SystemModelDecompose}
{\bf y}_c = {\bf y}_I + j{\bf y}_Q, &
\,\,\, {\bf x}_c = {\bf x}_I + j{\bf x}_Q, \nonumber \\
{\bf n}_c = {\bf n}_I+j{\bf n}_Q, &
\,\,\, \widetilde{{\bf H}}_c = \widetilde{{\bf H}}_I+j\widetilde{{\bf H}}_Q.
\end{eqnarray}
Further, we define
${\bf x}_r \in {\mathbb R}^{2k \times 1}$,
${\bf y}_r \in {\mathbb R}^{2N_rp \times 1}$,
${\bf H}_r \in {\mathbb R}^{2N_rp \times 2k}$,
and
${\bf n}_r \in {\mathbb R}^{2N_rp \times 1}$ as
\begin{eqnarray}
\label{SystemModelRealDef}
{\bf x}_r = [{\bf x}_I^T \hspace{2mm} {\bf x}_Q^T ]^T, &
\,\,\,\,\,
{\bf y}_r = [{\bf y}_I^T \hspace{2mm} {\bf y}_Q^T ]^T, \nonumber \\
{\bf H}_r = \left(\begin{array}{cc}\widetilde{{\bf H}}_I \hspace{2mm} -\widetilde{{\bf H}}_Q \\
\widetilde{{\bf H}}_Q  \hspace{5mm} \widetilde{{\bf H}}_I \end{array}\right), &
\,\,\,\,\,
{\bf n}_r = [{\bf n}_I^T \hspace{2mm} {\bf n}_Q^T ]^T.
\end{eqnarray}
Now, (\ref{SystemModelvec2}) can be written as
\begin{eqnarray}
\label{SystemModelReal}
{\bf y}_r & = & {\bf H}_r{\bf x}_r + {\bf n}_r.
\end{eqnarray}
Henceforth, we work with the real-valued system in
(\ref{SystemModelReal}). For notational simplicity, we drop
subscripts $r$ in (\ref{SystemModelReal}) and write
\begin{eqnarray}
\label{SystemModelII}
{\bf y} & = & {\bf H} {\bf x} + {\bf n},
\end{eqnarray}
where ${\bf H} = {\bf H}_r \in {\mathbb R}^{2N_rp \times 2k}$,
${\bf y} = {\bf y}_r \in {\mathbb R}^{2N_rp \times 1}$,
${\bf x} = {\bf x}_r \in {\mathbb R}^{2k \times 1}$, and
${\bf n} = {\bf n}_r \in {\mathbb R}^{2N_rp \times 1}$.
The channel coefficients are assumed to be known only at the
receiver but not at the transmitter.
Let $\mathbb A_i$ denote the ${\mathcal M}$-PAM signal set from which
$x_i$ ($i$th
entry of ${\bf x}$) takes values, $i=1,\cdots,2k$. Now, define a
$2k$-dimensional signal space $\mathbb S$ to be the Cartesian product
of $\mathbb A_1$ to $\mathbb A_{2k}$.  The ML solution is given by

\thanks{
\line(1,0){505}
\begin{equation*}
\hspace{1.3cm}
\left[
\begin{array}{ccccc}
\sum_{i=0}^{n-1}x_{0,i}\,t^i & \delta\sum_{i=0}^{n-1}x_{n-1,i}\,\omega_n^i\,t^i & \delta\sum_{i=0}^{n-1}x_{n-2,i}\,\omega_n^{2i}\,t^i & \cdots & \delta\sum_{i=0}^{n-1}x_{1,i}\,\omega_n^{(n-1)i}\,t^i \\
\sum_{i=0}^{n-1}x_{1,i}\,t^i & \sum_{i=0}^{n-1}x_{0,i}\,\omega_n^i\,t^i & \delta\sum_{i=0}^{n-1}x_{n-1,i}\,\omega_n^{2i}\,t^i & \cdots & \delta\sum_{i=0}^{n-1}x_{2,i}\,\omega_n^{(n-1)i}\,t^i \\
\sum_{i=0}^{n-1}x_{2,i}\,t^i & \sum_{i=0}^{n-1}x_{1,i}\,\omega_n^i\,t^i & \sum_{i=0}^{n-1}x_{0,i}\,\omega_n^{2i}\,t^i & \cdots & \delta\sum_{i=0}^{n-1}x_{3,i}\,\omega_n^{(n-1)i}\,t^i \\
\vdots & \vdots & \vdots & \vdots & \vdots \\
\sum_{i=0}^{n-1}x_{n-2,i}\,t^i & \sum_{i=0}^{n-1}x_{n-3,i}\,\omega_n^i\,t^i & \sum_{i=0}^{n-1}x_{n-4,i}\,\omega_n^{2i}\,t^i & \cdots & \delta \sum_{i=0}^{n-1}x_{n-1,i}\,\omega_n^{(n-1)i}t^i \\
\sum_{i=0}^{n-1}x_{n-1,i}\,t^i & \sum_{i=0}^{n-1}x_{n-2,i}\,\omega_n^i\,t^i & \sum_{i=0}^{n-1}x_{n-3,i}\,\omega_n^{2i}\,t^i & \cdots & \sum_{i=0}^{n-1}x_{0,i}\,\omega_n^{(n-1)i}\,t^i
\end{array}
\right]. \hspace{10mm} (\mbox{11.a})
\label{stbc}
\end{equation*}
}

\begin{eqnarray}
\label{MLdetection}
{\bf d}_{ML} & = & {\mbox{arg min}\atop{{\bf d} \in {\mathbb S}}} \
 \Vert {\bf y} - {\bf H}{\bf d} \Vert ^2 \nonumber \\
& = & {\mbox{arg min}\atop{{\bf d} \in {\mathbb S}}} \
{\bf d}^T {\bf H}^T{\bf H}{\bf d} - 2{\bf y}^T{\bf H}{\bf d},
\end{eqnarray}
whose complexity is exponential in $k$ \cite{verdu}. 

\subsection{High-rate Non-orthogonal STBCs from CDA}
\label{sec21}
We focus on the detection of square (i.e.,
$n\hspace{-0.5mm}=\hspace{-0.5mm}p\hspace{-0.5mm}=\hspace{-0.5mm}N_t$),
full-rate (i.e.,
$k\hspace{-0.5mm}=\hspace{-0.5mm}pn\hspace{-0.5mm}=\hspace{-0.5mm}N_t^2$),
circulant (where the weight matrices ${\bf A}_c^{(i)}$'s are
permutation type), non-orthogonal STBCs from CDA \cite{gcom08}, whose
construction for arbitrary number of transmit antennas $n$ is given by
the matrix in (11.a) given at the bottom of this page \cite{bsr}:

In (11.a), {\small $\omega_n=e^{\frac{{\bf j}2\pi}{n}}$,
${\bf j}=\sqrt{-1}$, and $x_{u,v}$, $0\leq u,v \leq n-1$} are the
data symbols from a QAM alphabet. When $\delta=e^{\sqrt{5}\,{\bf j}}$
and $t=e^{{\bf j}}$, the STBC in (11.a) achieves full transmit
diversity (under ML decoding) as well as information-losslessness \cite{bsr}.
When $\delta=t=1$, the code ceases to be of full-diversity (FD), but
continues to be information-lossless (ILL) \cite{hassibi2},\cite{cpa}. 
High spectral
efficiencies with large $n$ can be achieved using this code construction.
For example, with $n=32$ transmit antennas, the $32\times 32$ STBC from
(11.a) with {\small 16-QAM} and rate-3/4 turbo code achieves a spectral
efficiency of 96 bps/Hz. This high spectral efficiency is achieved along
with the full-diversity of order $nN_r$. However, since these STBCs are
non-orthogonal, ML detection gets increasingly impractical for large $n$.
Consequently, a key challenge in realizing the benefits of these large
STBCs in practice is that of achieving near-ML performance for large $n$
at low detection complexities.
Our proposed detector, termed as the {\em multistage likelihood ascent
search ($M$-LAS) detector}, presented in the following section essentially
addresses this challenging issue.

\section{Proposed Multistage LAS Detector }
\label{sec3}
The proposed $M$-LAS algorithm consists of a sequence of likelihood-ascent
search stages, where the likelihood increases monotonically with every
search stage. Each search stage consists of several sub-stages. There
can be at most $M$ sub-stages, each consisting of one or more iterations
(the first sub-stage can have one or more iterations, whereas all the
other sub-stages can have at most one iteration). In the first sub-stage,
the algorithm updates one symbol per iteration such that the likelihood
monotonically increases from one iteration to the next until a local
minima is reached. Upon reaching this local minima, the algorithm
initiates the second sub-stage. 
\newpage
In the second sub-stage, a 2-symbol
update is tried to further increase the likelihood. If the algorithm
succeeds in increasing the likelihood by 2-symbol update, it starts
the next search stage. If the algorithm does not succeed in the second
sub-stage, it goes to the third sub-stage where a 3-symbol update is
tried to further increase the likelihood. Essentially, in the $K$th
sub-stage, a $K$-symbol update is tried to further increase the
likelihood. This goes on until $a)$ either the algorithm succeeds in
the $K$th sub-stage for some $K \leq M$ (in which case a new search
stage is initiated), or $b)$ the algorithm terminates.

The $M$-LAS algorithm starts with an initial solution ${\bf d}^{(0)}$,
given by ${\bf d}^{(0)} = {\bf B}{\bf y}$, where ${\bf B}$ is
the initial solution filter, which can be a matched filter (MF) or
zero-forcing (ZF) filter or MMSE filter. The index $m$ in ${\bf d}^{(m)}$
denotes the iteration number in a sub-stage of a given search stage. The
ML cost function after the $k$th iteration in a given search stage is 
\begin{eqnarray}
\label{Ck}
C^{(k)} & = & {\bf d}^{(k)^T} {\bf H}^T{\bf H} {\bf d}^{(k)}-2{\bf y}^T{\bf H}{\bf d}^{(k)}.
\end{eqnarray}

\subsection{One-symbol Update}
\label{sec_1symb}
Let us assume that we update the $p$th symbol in the $(k+1)$th iteration;
$p$ can take value from $1,\cdots,N_t$ for ${\mathcal M}$-PAM and
$1,\cdots,2N_t$ for ${\mathcal M}$-QAM. The update rule can be written as
\begin{eqnarray}
\label{UpdateK}
{\bf d}^{(k+1)} & = & {\bf d}^{(k)} + \lambda_p^{(k)} {\bf e}_p,
\end{eqnarray}
where ${\bf e}_p$ denotes the unit vector with its $p$th entry only as one,
and all other entries as zero. Also, for any iteration $k$, ${\bf d}^{(k)}$
should belong to the space $\mathbb S$, and therefore $\lambda_p^{(k)}$ can
take only certain integer values. For example, in case of 4-PAM or 16-QAM
$\big($both have the same signal set ${\mathbb A}_p = \{ -3, -1, 1, 3\}\big)$,
$\lambda_p^{(k)}$ can take values only from $\{-6,-4,-2,0,2,4,6\}$. Using
(\ref{Ck}) and (\ref{UpdateK}), and defining a matrix ${\bf G}$ as
\begin{eqnarray}
{\bf G} & \Define & {\bf H}^{T}{\bf H},
\end{eqnarray}
we can write the cost difference as
\begin{eqnarray}
\label{Ck+1MinusCk}
\hspace{-7mm}
\Delta C_p^{k+1} & \Define & C^{(k+1)} - C^{(k)} \nonumber \\
& =  & \lambda_p^{(k)^2} ({\bf G})_{p,p}-2\lambda_p^{(k)} z^{(k)}_p,
\end{eqnarray}
where
${\bf h}_p$ is the $p$th column of ${\bf H}$,
{\small ${\bf z}^{(k)}={\bf H}^T({\bf y}-{\bf H}{\bf d}^{(k)})$},
$z^{(k)}_p$ is the $p$th entry of the ${\bf z}^{(k)}$ vector, and
$\left({\bf G}\right)_{p,p}$ is the $(p,p)$th
entry of the ${\bf G}$
matrix. Also, let us define $a_p$ and $l_p^{(k)}$ as
\begin{eqnarray}
\label{algodefines}
a_p \,\,\, = \,\,\, ({\bf G})_{p,p}\,, \quad \quad l_p^{(k)} \,\,\, = \,\,\, \vert \lambda_p^{(k)} \vert.
\end{eqnarray}
With the above variables defined, we can rewrite (\ref{Ck+1MinusCk}) as
\begin{eqnarray}
\label{CostDiff}
\hspace{-8mm}
\Delta C_p^{k+1} & = & l_p^{(k)^2}a_p  - 2l_p^{(k)} \vert z_p^{(k)} \vert \, \mbox{sgn}(\lambda_p^{(k)}) \, \mbox{sgn}(z_p^{(k)}),
\end{eqnarray}
where $\mbox{sgn}(.)$ denotes the signum function. For the ML cost function
to reduce from the $k$th to the $(k+1)$th iteration, the cost difference
should be negative. Using this fact and that $a_p$ and $l_p^{(k)}$ are
non-negative quantities, we can conclude from (\ref{CostDiff}) that
the sign of $\lambda_p^{(k)}$ must satisfy
\begin{eqnarray}
\label{conc1}
\mbox{sgn}(\lambda_p^{(k)}) & = & \mbox{sgn}(z_p^{(k)}).
\end{eqnarray}
Using (\ref{conc1}) in (\ref{CostDiff}), the ML cost difference can be
rewritten as
\begin{eqnarray}
\label{CostDiff2}
\mathcal F(l_p^{(k)}) & \Define & \Delta C_p^{k+1}
\,\,\, = \,\,\,\, l_p^{(k)^2}a_p  - 2l_p^{(k)} \vert z_p^{(k)} \vert.
\end{eqnarray}
For {\small $\mathcal F(l_p^{(k)})$} to be non-positive, the necessary and
sufficient condition from (\ref{CostDiff2}) is that
\begin{eqnarray}
\label{NSC1}
l_p^{(k)} & < & \frac {2 \vert z_p^{(k)} \vert} {a_p}.
\end{eqnarray}
However, we can find the value of $l_p^{(k)}$ which satisfies (\ref{NSC1})
and at the same time gives the largest descent in the ML cost function from
the $k$th to the $(k+1)$th iteration (when symbol $p$ is updated).
Also, $l_p^{(k)}$ is constrained to take only certain integer values, and
therefore the brute-force way to get optimum $l_p^{(k)}$ is to evaluate
{\small $\mathcal F(l_p^{(k)})$} at all possible values of $l_p^{(k)}$.
This would become computationally expensive as the constellation
size ${\mathcal M}$ increases. However, for the case of 1-symbol update,
we could obtain a closed-form expression for the optimum $l_p^{(k)}$
that minimizes {\small $\mathcal F(l_p^{(k)})$}, which is given by
(corresponding theorem and proof are given in the Appendix)
\begin{eqnarray}
\label{Optdp}
l_{p,opt}^{(k)} & = & 2 \left\lfloor \frac { \vert z_p^{(k)} \vert } { 2 a_p} \right\rceil,
\end{eqnarray}
where $\lfloor . \rceil$ denotes the rounding operation, where for
a real number $x$, $\lfloor x \rceil$ is the integer closest to $x$.
If the $p$th symbol in ${\bf d}^{(k)}$, i.e., $d_p^{(k)}$, were indeed
updated, then the new value of the symbol would be given by
\begin{eqnarray}
\label{NewDpk}
\tilde{d}_p^{(k+1)} &=& d_p^{(k)}+l_p^{(k)} \mbox{sgn}(z_p^{(k)}).
\end{eqnarray}
However, $\tilde{d}_p^{(k+1)}$ can take values only in the set
${\mathbb A_p}$, and therefore we need to check for the possibility of
$\tilde{d}_p^{(k+1)}$ being greater than $({\mathcal M}-1)$ or less than
$-({\mathcal M}-1)$.
If $\tilde{d}_p^{(k+1)} > ({\mathcal M}-1)$, then $l_p^{(k)}$ is adjusted
so that the
new value of $\tilde{d}_p^{(k+1)}$ with the adjusted value of $l_p^{(k)}$
using (\ref{NewDpk}) is $({\mathcal M}-1)$. Similarly, if
$\tilde{d}_p^{(k+1)} < -({\mathcal M}-1)$, then $l_p^{(k)}$ is adjusted
so that the
new value of $\tilde{d}_p^{(k+1)}$ is $-({\mathcal M}-1)$. Let
$\tilde{l}_{p,opt}^{(k)}$ be obtained from $l_{p,opt}^{(k)}$ after these
adjustments. It can be shown that if {\small $\mathcal F(l_{p,opt}^{(k)})$}
is non-positive, then {\small $\mathcal F(\tilde{l}_{p,opt}^{(k)})$} is
also non-positive. We compute {\small $\mathcal F(\tilde{l}_{p,opt}^{(k)})$},
$\forall$ $ p=1,\cdots,2N_t^2$. Now, let
\begin{eqnarray}
s & = & {\mbox{arg min}\atop p} \,\,\,\mathcal F(\tilde{l}_{p,opt}^{(k)}).
\end{eqnarray}
If {\small $\mathcal F(\tilde{l}_{s,opt}^{(k)}) < 0$}, the update for the
$(k+1)$th iteration is
\begin{eqnarray}
\label{AlgoUpdateK}
{\bf d}^{(k+1)} &=& {\bf d}^{(k)} + \tilde{l}_{s,opt}^{(k)} \, \mbox{sgn}(z_s^{(k)})\,{\bf e}_s, \\{\bf z}^{(k+1)} &=& {\bf z}^{(k)} - \tilde{l}_{s,opt}^{(k)} \,\mbox{sgn}(z_s^{(k)})\,{\bf g}_s,
\label{eqx}
\end{eqnarray}
where ${\bf g}_s$ is the $s$th column of ${\bf G}$. The update in (\ref{eqx})
follows from the definition of ${\bf z}^{(k)}$ in (\ref{Ck+1MinusCk}).
If $\mathcal F(\tilde{l}_{s,opt}^{(k)}) \geq 0$, then the 1-symbol update
search terminates. The data vector at this point is referred to as
`1-symbol update local minima.' After reaching the 1-symbol update local
minima, we look for a further decrease in the cost function by updating
multiple symbols simultaneously.

\subsection{Why Multiple Symbol Updates?}
\label{sec_why}
The motivation for trying out multiple symbol updates can be explained
as follows. Let ${\mathbb L}_K \subseteq {\mathbb S}$ denote the set of
data vectors such that for any ${\bf d} \in {\mathbb L}_K$, if a $K$-symbol
update is performed on ${\bf d}$ resulting in a vector ${\bf d}'$, then
$||{\bf y}-{\bf H}{\bf d}'|| \geq  ||{\bf y}-{\bf H}{\bf d}||$. We
note that ${\bf d}_{ML} \in {\mathbb L}_K, \forall \, K=1,2,\cdots,2N_t$,
because any number of symbol updates on ${\bf d}_{ML}$ will not decrease
the cost function. We define another set
${\mathbb M}_K = \bigcap_{j=1}^{K} {\mathbb L}_j$. Note that
${\bf d}_{ML} \in {\mathbb M}_K, \forall \, K=1,2,\cdots,2N_t$, and
${\mathbb M}_{2N_t} = \{{\bf d}_{ML}\}$, i.e.,
${\mathbb M}_{2N_t}$ is a singleton set with  ${\bf d}_{ML}$ as the
only element.
It is noted that if the updates are done optimally, then the output
of the $K$-LAS algorithm converges to a vector in ${\mathbb M}_K$.
Also, $|{\mathbb M}_{K+1}| \leq |{\mathbb M}_{K}|,
\, K=1,2,\cdots,2N_t-1$.
For any ${\bf d} \in {\mathbb M}_{K}$, $K=1,2,\cdots,2N_t$ and
${\bf d} \neq {\bf d}_{ML}$, it can be seen that ${\bf d}$ and
${\bf d}_{ML}$ will differ in $K+1$ or more locations.
The probability that ${\bf d}_{ML} = {\bf x}$ increases with increasing
SNR, and so the separation between ${\bf d} \in \mathbb{M}_K$ and
${\bf x}$ will monotonically increase with increasing $K$.
Since ${\bf d}_{ML} \in {\mathbb M}_{K}$, and $|{\mathbb M}_{K}|$
decreases monotonically with increasing $K$, there will be lesser
non-ML data vectors to which the algorithm can converge to for
increasing $K$.
Therefore, the probability of
the noise vector ${\bf n}$ inducing an error would decrease with increasing
$K$. This indicates that $K$-symbol updates with large $K$ could get near
to ML performance with increasing complexity for increasing $K$.

\subsection{$K$-symbol Update, $1<K\leq 2N_t^2$}
\label{sec_Ksymb}
In this subsection, we present the update algorithm for the general case
where $K$ symbols, $1< K \leq 2N_t^2$, are updated simultaneously in one
iteration. $K$-symbol updates can be done in $2N_t^2\choose K$ ways, among
which we seek to find that update which gives the largest reduction in
the ML cost. Assume that in the $(k+1)$th iteration, $K$ symbols at the
indices $i_1,i_2,\cdots,i_K$ of ${\bf d}^{(k)}$ are updated. Each $i_j$,
$j=1,2,\cdots,K$, can take values from $1,2,\cdots,N_t^2$ for
${\mathcal M}$-PAM and $1,2,\cdots,2N_t^2$ for ${\mathcal M}$-QAM.
Further, define the set of indices,
${\mathcal U} \, \Define \, \{ i_1,i_2,\cdots,i_K \}$. The update rule for
the $K$-symbol update can then be written as
\begin{eqnarray}
\label{UpdateKK}
{\bf d}^{(k+1)} & = & {\bf d}^{(k)} + \sum_{j=1}^{K} \lambda_{i_j}^{(k)} {\bf e}_{i_j}.
\end{eqnarray}
For any iteration $k$, ${\bf d}^{(k)}$ belongs to the space $\mathbb S$,
and therefore $\lambda_{i_j}^{(k)}$ can take only certain integer values.
In particular, $\lambda_{i_j}^{(k)}\in {\mathbb A}_{i_j}^{(k)}$, where
${\mathbb A}_{i_j}^{(k)}\Define \{ x\vert (x + d_{i_j}^{(k)})\in {\mathbb A}_{i_j}, x \ne 0\}$.
For example,
for 16-QAM, ${\mathbb A}_{i_j}=\{-3,-1,1,3\}$, and if $d^{(k)}_{i_j}$ is -1,
then ${\mathbb A}_{i_j}^{(k)}$ = $\{-2,2,4\}$.
Using (\ref{Ck}), we can
write the cost difference function
${\small \Delta C^{k+1}_{\mathcal U}(\lambda^{(k)}_{i_1}, \lambda^{(k)}_{i_2}, \cdots ,\lambda^{(k)}_{i_K})} \, \Define \, C^{(k+1)} - C^{(k)}$ as
\begin{eqnarray}
\label{eqpqK}
\hspace{-9mm}
\Delta C^{k+1}_{\mathcal U}(\lambda^{(k)}_{i_1},\lambda^{(k)}_{i_2},\cdots,\lambda^{(k)}_{i_K}) & \hspace{-1mm} = & \hspace{-1mm} \sum_{j=1}^{K} \lambda_{i_j}^{(k)^2} ({\bf G})_{i_j,i_j} \nonumber \\
&\hspace{-49mm} & \hspace{-42mm} 
+ \,\, 2 \sum_{q=1}^{K} \sum_{p=q+1}^{K} \, \lambda_{i_p}^{(k)}\lambda_{i_q}^{(k)}({\bf G})_{{i_p},{i_q}} - \, 2 \sum_{j=1}^{K} \lambda_{i_j}^{(k)} z^{(k)}_{i_j},
\end{eqnarray}
where {\small $\lambda_{i_j}^{(k)} \in {\mathbb A}_{i_j}^{(k)}$}, which
can be compactly written as
{\small $(\lambda^{(k)}_{i_1},\lambda^{(k)}_{i_2},\cdots,\lambda^{(k)}_{i_K}) \in {\mathbb A}_{\mathcal U}^{(k)}$},
where ${\mathbb A}_{\mathcal U}^{(k)}$ denotes the Cartesian product of
${\mathbb A}_{i_1}^{(k)}$, ${\mathbb A}_{i_2}^{(k)}$ through to ${\mathbb A}_{i_K}^{(k)}$.

For a given ${\mathcal U}$, in order to decrease the ML cost, we would like
to choose the value of the $K$-tuple
{\small $(\lambda_{i_1}^{(k)},\lambda_{i_2}^{(k)},\cdots,\lambda_{i_K}^{(k)})$}
such that the cost difference given by (\ref{eqpqK}) is negative. If multiple
$K$-tuples exist for which the cost difference is negative, we choose the
$K$-tuple which gives the most negative cost difference.

Unlike for 1-symbol update, for $K$-symbol update we do not have a
closed-form expression for
{\small $(\lambda_{{i_1},opt}^{(k)},\lambda_{{i_2},opt}^{(k)},\cdots,\lambda_{{i_K},opt}^{(k)})$} which minimizes
the cost difference over ${\mathbb A}_{\mathcal U}^{(k)}$,
since the cost difference
is a function of $K$ discrete valued variables. Consequently, a
brute-force method is to evaluate
{\small $ \Delta C^{k+1}_{\mathcal U}(\lambda_{i_1}^{(k)},\lambda_{i_2}^{(k)},\cdots,\lambda_{i_K}^{(k)})$}
over all possible values of
$(\lambda_{i_1}^{(k)},\lambda_{i_2}^{(k)},\cdots,\lambda_{i_K}^{(k)})$.
Approximate methods can be adopted to solve this problem using lesser
complexity. One method based on zero-forcing is as follows. The
cost difference function in (\ref{eqpqK}) can be rewritten as
\begin{eqnarray}
\label{SelectKTwoSubOptK}
\hspace{-6mm}
\Delta C^{k+1}_{\mathcal U}(\lambda_{i_1}^{(k)},\lambda_{i_2}^{(k)},\cdots,\lambda_{i_K}^{(k)}) & = & {\bf \Lambda}_{\mathcal U}^{(k)^T} {\bf F}_{\mathcal U} \,  {\bf \Lambda}_{\mathcal U}^{(k)} \nonumber \\
& & -\,\, 2{\bf \Lambda}_{\mathcal U}^{(k)^T}{\bf z}_{\mathcal U}^{(k)},
\end{eqnarray}
where
{\small ${\bf \Lambda}_{\mathcal U}^{(k)}\Define [\lambda_{i_1}^{(k)} \lambda_{i_2}^{(k)}\cdots \lambda_{i_K}^{(k)}]^T$},
{\small ${\bf z}_{\mathcal U}^{(k)} \Define [ {z}_{i_1}^{(k)} {z}_{i_2}^{(k)}\cdots {z}_{i_K}^{(k)}]^T$}, and
${\bf F}_{\mathcal U} \in {\mathbb R}^{K \times K}$, where
{\small ${({\bf F}_{\mathcal U})}_{p,q} = ({\bf G})_{{i_p},{i_q}}$} and
$p,q \in \{1,2,\cdots,K\}$.
Since $ \Delta C^{k+1}_{\mathcal U}(\lambda_{i_1}^{(k)},\lambda_{i_2}^{(k)},\cdots,\lambda_{i_K}^{(k)})$ is a
strictly convex quadratic function of ${\bf \Lambda}_{\mathcal U}^{(k)}$ 
(the Hessian ${\bf F}_{\mathcal U}$ is positive definite with probability 1),
a unique global minima exists, and is given by
\begin{eqnarray}
\label{SelectKTwoSubOptSolnRK}
{\tilde{\bf \Lambda}}_{\mathcal U}^{(k)} & = & {\bf F}_{\mathcal U}^{-1}\,\, {\bf z}_{\mathcal U}^{(k)}.
\end{eqnarray}
However, the solution given by (\ref{SelectKTwoSubOptSolnRK}) need not lie
in ${\mathbb A}_{\mathcal U}^{(k)}$. So, we
first round-off the solution as
\begin{eqnarray}
\label{SelectKTwoSubOptSolnRndK}
{\widehat{\bf \Lambda}}_{\mathcal U}^{(k)} & = & 2\Big\lfloor 0.5{\tilde{\bf \Lambda}}_{\mathcal U}^{(k)} \Big\rceil,
\end{eqnarray}
where the operation in (\ref{SelectKTwoSubOptSolnRndK}) is done element-wise,
since ${\tilde{\bf \Lambda}}_{\mathcal U}^{(k)}$ is a vector. Further, let
{\small ${\widehat{\bf \Lambda}}_{\mathcal U}^{(k)} \Define [ \widehat{\bf \lambda}_{i_1}^{(k)}  \widehat{\bf \lambda}_{i_2}^{(k)}  \cdots  \widehat{\bf \lambda}_{i_K}^{(k)}  ]^T$}. It is still possible that the solution
${\widehat{\bf \Lambda}}_{\mathcal U}^{(k)}$ in (\ref{SelectKTwoSubOptSolnRndK})
need not lie in ${\mathbb A}_{\mathcal U}^{(k)}$.
This would result in $d_{i_j}^{(k+1)} \notin {\mathbb A}_{i_j}$ for some $j$.
For example, if ${\mathbb A}_{i_j}$ is ${\mathcal M}$-PAM, then
$d_{i_j}^{(k+1)} \notin {\mathbb A}_{i_j}$ if
{\small $d_{i_j}^{(k)} + \widehat{\bf \lambda}_{i_j}^{(k)}  > ({\mathcal M}-1)$} or {\small $d_{i_j}^{(k)} + \widehat{\bf \lambda}_{i_j}^{(k)}  < -({\mathcal M}-1)$}  .
In such cases, we propose the following adjustment to
$\widehat{\bf \lambda}_{i_j}^{(k)}$ for $j=1,2,\cdots,K$:
{\small
\begin{eqnarray}
\hspace{-7mm}
\widehat{\lambda}_{i_j}^{(k)} & \hspace{-2.5mm} = & \hspace{-2.5mm} \left\{
\begin{array}{ll}
({\mathcal M}-1)-d_{i_j}^{(k)}, & \hspace{-3mm} \mbox{when} \,\, \widehat{\lambda}_{i_j}^{(k)} + d_{i_j}^{(k)} > ({\mathcal M}-1) \\
-({\mathcal M}-1)-d_{i_j}^{(k)}, & \hspace{-3mm} \mbox{when} \,\, \widehat{\lambda}_{i_j}^{(k)} + d_{i_j}^{(k)} < -({\mathcal M}-1).
\end{array}\right. \hspace{-4mm}
\label{new1x}
\end{eqnarray}
}
After these adjustments, we are
guaranteed that
{\small ${\widehat{\Lambda}}_{\mathcal U}^{(k)} \in {\mathbb A}_{\mathcal U}^{(k)}$.}
Therefore, the new cost difference function value is given by
$\Delta C^{k+1}_{\mathcal U}({\widehat{\lambda}}_{i_1}^{(k)},{\widehat{\lambda}}_{i_2}^{(k)},\cdots,{\widehat{\lambda}}_{i_K}^{(k)})$.
It is noted that the complexity of this approximate method does not depend on
the size of the set ${\mathbb A}_{\mathcal U}^{(k)}$, i.e., it has constant
complexity. Through simulations, we have observed that this
approximation results in a performance close to that of the brute-force
method for $K=2$ and 3.
Defining the optimum ${\mathcal U}$ for the approximate method as
$\hat{{\mathcal U}}$,
we can write
\begin{eqnarray}
\label{SelectpqTwoNotBruteK}
\hat{{\mathcal U}}
& \Define & ({\hat i}_1,{\hat i}_2,\cdots,{\hat i}_K) \nonumber \\
& = & {\mbox{arg min} \atop {\mathcal U}} \,\, \Delta C^{k+1}_{\mathcal U}(  {\widehat {\lambda}}_{i_1}^{(k)}, {\widehat{\lambda}}_{i_2}^{(k)},\cdots,{\widehat{\lambda}}_{i_K}^{(k)}).
\end{eqnarray}
The $K$-update is successful and the update is done only if
{\small $ \Delta C^{k+1}_{\hat {\mathcal U}}({\widehat{\lambda}}_{\hat{i_1}}^{(k)},{\widehat{\lambda}}_{\hat{i_2}}^{(k)},\cdots,{\widehat{\lambda}}_{\hat{i_K}}^{(k)}) < 0 $}.
The update rules for the ${\bf z}^{(k)}$ and ${\bf d}^{(k)}$ vectors are
given by
\begin{eqnarray}
{\bf z}^{(k+1)} & = & {\bf z}^{(k)} - \sum_{j=1}^{K} \, \widehat{\lambda}_{\hat{i_j}}^{(k)} \, {\bf g}_{\hat{i_j}} \, , \\
{\bf d}^{(k+1)} & = & {\bf d}^{(k)} + \sum_{j=1}^{K} \, \widehat{\lambda}_{\hat{i_j}}^{(k)} \, {\bf e}_{\hat{i_j}} \, .
\end{eqnarray}

\subsection{Computational Complexity of the $M$-LAS Algorithm}
\label{sec_comp}
\begin{figure}
\hspace{-4mm}
\includegraphics[width=3.60in]{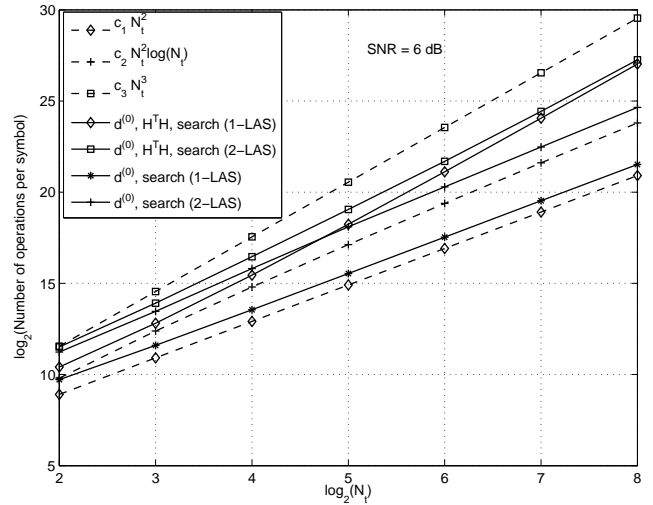}
\caption{
Computational complexity of the proposed $M$-LAS algorithm in
decoding non-orthogonal STBCs from CDA. MMSE initial vector,
4-QAM, SNR = 6 dB.
}
\label{fig1}
\vspace{-2mm}
\end{figure}

The complexity of the proposed $M$-LAS algorithm comprises of three components,
namely, $i)$ computation of the initial vector ${\bf d}^{(0)}$, $ii)$
computation of ${\bf H}^T{\bf H}$, and $iii)$ the search operation. Figure
\ref{fig1} shows the per-symbol complexity plots as a function of $N_t=N_r$
for 4-QAM at an SNR of 6 dB using MMSE initial vector. Two good properties
of the STBCs from CDA are useful in achieving low orders of
complexity for the computation of ${\bf d}^{(0)}$ and ${\bf H}^T{\bf H}$.
They are: $i)$ the weight matrices ${\bf A}_c^{(i)}$'s are {\em permutation
type}, and $ii)$ the $N_t^2\times N_t^2$ matrix formed with
$N_t^2\times 1$-sized ${\bf a}_c^{(i)}$ vectors as columns is a {\em scaled
unitary matrix}. These properties allow the computation of MMSE/ZF initial
solution in $O(N_t^3 N_r)$ complexity, i.e., in $O(N_tN_r)$ per-symbol
complexity since there are $N_t^2$ symbols in one STBC matrix. Likewise,
the computation of ${\bf H}^T{\bf H}$ can be done in $O(N_t^3)$ per-symbol
complexity.

The average per-symbol complexities of the 1-LAS and 2-LAS search operations
are $O(N_t^2)$ and $O(N_t^2\log N_t)$, respectively, which can be explained
as follows. The average search complexity is the complexity of one search
stage times the mean number of search stages till the algorithm terminates.
For 1-LAS, the number of search stages is always one. There are multiple
iterations in the search, and in each iteration all possible
$2N_t^2 \choose 1$ 1-symbol updates are considered. So, the per-iteration
complexity in 1-LAS is $O(N_t^2)$, i.e., $O(1)$ complexity per symbol.
Further, the mean number of iterations before the algorithm terminates in
1-LAS was found to be $O(N_t^2)$ through simulations. So, the overall
per-symbol complexity of 1-LAS is $O(N_t^2)$. In 2-LAS, the complexity of
the 2-symbol update dominates over the 1-symbol update. Since there are
$2N_t^2 \choose 2$ possible 2-symbol updates, the complexity of one search
stage is $O(N_t^4)$, i.e., $O(N_t^2)$ complexity per symbol. The mean
number of stages till the algorithm terminates in 2-LAS was found to be
$O(\log N_t)$ through simulations. Therefore, the overall per-symbol
complexity of 2-LAS is $O(N_t^2 \log N_t)$. These can be observed from
Fig. \ref{fig1}, where it can be seen that the per-symbol complexity in
the initial vector computation plus the 1-LAS/2-LAS search operation is
$O(N_t^2)$/$O(N_t^2 \log N_t)$; i.e., 1-LAS and 2-LAS complexity plots
run parallel to the $c_1N_t^2$ and $c_2N_t^2 \log N_t$ lines, respectively.
With the computation of ${\bf H}^T{\bf H}$ included, the complexity order
is more than $N_t^2$. From the slopes of the plots in Fig. \ref{fig1}, we
find that the overall complexities for $N_t=16$ and 32 are proportional to
$N_t^{2.5}$ and $N_t^{2.7}$, respectively.

For the special case of ILL-only STBCs (i.e., $\delta=t=1$), the
complexity involved in computing ${\bf d}^{(0)}$ and ${\bf H}^T{\bf H}$
can be reduced further. This becomes possible due to the following
property of ILL-only STBCs. 
Let ${\bf V}_{a}$ be the complex $N_t^2\times N_t^2$ matrix with
${\bf a}_c^{(i)}$ as its $i$th column. The computation of ${\bf d}^{(0)}$
(or ${\bf H}^T{\bf H}$) involves multiplication of ${\bf V}_a^H$ with another
vector (or matrix). The columns of ${\bf V}_a^H$ can be permuted in such a
way that the permuted matrix is block-diagonal, where each block is a
$N_t\times N_t$ DFT matrix for $\delta=t=1$. So, the multiplication of
${\bf V}_a^H$ by any vector becomes equivalent to a $N_t$-point DFT
operation, which can be efficiently computed using FFT in
$O(N_t \log N_t)$ complexity. Using this simplification, the per-symbol
complexity  of computing ${\bf H}^T{\bf H}$ is reduced from $O(N_t^3)$ to
$O(N_t^2\log N_t)$.
Computing ${\bf d}^{(0)}$
using MMSE filter involves the computation of
$\frac{1}{N_t}{\bf V}_a^H({\bf I}\otimes (({\bf H}_c^H{\bf H}_c + \frac{1}{\gamma N_t}{\bf I})^{-1}{\bf H}_c^H)){\bf y}_c$. The complexity of computing
the vector
$({\bf I}\otimes (({\bf H}_c^H{\bf H}_c + \frac{1}{\gamma N_t}{\bf I})^{-1}{\bf H}_c^H)){\bf y}_c$ is $O(N_t^2N_r)$,
and the complexity of computing
${\bf V}_a^H({\bf I}\otimes (({\bf H}_c^H{\bf H}_c + \frac{1}{\gamma N_t}{\bf I})^{-1}{\bf H}_c^H)){\bf y}_c$ is $O(N_t^3N_r)$.
In the case of ILL-only STBC, because of the above-mentioned property,
the complexity of computing
${\bf V}_a^H({\bf I}\otimes (({\bf H}_c^H{\bf H}_c + \frac{1}{\gamma N_t}{\bf I})^{-1}{\bf H}_c^H)){\bf y}_c$
gets reduced to $O(N_t^2\log N_t)$ from $O(N_t^3N_r)$. So the total
complexity for computing ${\bf d}^{(0)}$ in ILL-only STBC is
$O(N_t^2N_r)+O(N_t^2\log N_t)$, which gives a per-symbol complexity
of $O(N_r)+O(\log N_t)$.
So, the overall per-symbol complexity for 1-LAS detection of ILL-STBCs
is $O(N_t^2 \log N_t)$.

\subsection{Generation of Soft Outputs}
\label{sec_soft}
We propose to generate soft values at the $M$-LAS output for all the
individual bits that constitute the ${\mathcal M}$-PAM/${\mathcal M}$-QAM
symbols as follows. These output values are fed as soft inputs to the
decoder in a coded system. Let
{\small ${\bf d}=[\widehat{x}_1,\widehat{x}_2,\cdots,\widehat{x}_{2N_t^2}]$},
$\widehat{x}_{i} \in \mathbb{A}_i$ denote the detected output symbol vector
from the $M$-LAS algorithm. Let the symbol $\widehat{x}_i$ map to the bit
vector {\small ${\bf b}_i = [b_{i,1},b_{i,2},\cdots,b_{i,K_i}]^T$}, where
{\small $K_i = \log_2|\mathbb{A}_i|$}, and {\small $b_{i,j} \in \{+1,-1\}$},
{\small $i=1,2,\cdots,2N_t^2$} and {\small $j=1,2,\cdots,K_i$}. Let
$\tilde{b}_{i,j} \in \mathbb{R}$ denote the soft value for the
$j$th bit of the $i$th symbol. Given ${\bf d}$, we need to find
$\tilde{b}_{i,j}, \, \forall \, (i,j)$.

Note that the quantity $\Vert{\bf y}-{\bf H}{\bf d}\Vert^2$ is inversely
related to the likelihood that ${\bf d}$ is indeed the transmitted
symbol vector. Let the ${\bf d}$ vector with its $j$th bit of the $i$th
symbol forced to +1 be denoted as vector ${\bf d}_i^{j+}$. Likewise, let
${\bf d}_i^{j-}$ be the vector ${\bf d}$ with its $j$th bit of the $i$th
symbol forced to -1. Then the quantities
$\Vert{\bf y}-{\bf H}{\bf d}_i^{j+}\Vert^2$ and
$\Vert{\bf y}-{\bf H}{\bf d}_i^{j-}\Vert^2$
are inversely related to the likelihoods that the $j$th bit of the
$i$th transmitted symbol is +1 and -1, respectively. So, if
$\Vert{\bf y}-{\bf H}{\bf d}_i^{j-}\Vert^2-\Vert{\bf y}-{\bf H}{\bf d}_i^{j+}\Vert^2$ is
+ve (or -ve), it indicates that the $j$th bit of the $i$th transmitted
symbol has a higher likelihood of being +1 (or -1). So, the quantity
$\Vert{\bf y}-{\bf H}{\bf d}_i^{j-}\Vert^2 - \Vert{\bf y}-{\bf H}{\bf d}_i^{j+}\Vert^2$,
appropriately normalized to avoid unbounded increase for increasing
$N_t$, can be a good soft value for the $j$th bit of the $i$th symbol.
With this motivation, we generate the soft output value for the $j$th bit
of the $i$th symbol as
\begin{eqnarray}
\tilde{b}_{i,j} & = & \frac { \Vert {\bf y} - {\bf H}{\bf d}_i^{j-} \Vert^2 - \Vert {\bf y} - {\bf H}{\bf d}_i^{j+} \Vert^2 } {\Vert {\bf h}_i \Vert ^2},
\label{soft}
\end{eqnarray}
where the normalization by $\Vert {\bf h}_i \Vert^2$ is to contain
unbounded increase of $\tilde{b}_{i,j}$ for increasing $N_t$.
The RHS in the above can be efficiently computed in terms of
${\bf z}$ and ${\bf G}$ as follows.
Since
{\small ${\bf d}_i^{j+}$} and {\small ${\bf d}_i^{j-}$} differ only in
the $i$th entry, we can write
\begin{eqnarray}
\label{cd1}
{\bf d}_i^{j-} & = & {\bf d}_i^{j+} + \lambda_{i,j} {\bf e}_i.
\end{eqnarray}
Since we know ${\bf d}_i^{j-}$ and ${\bf d}_i^{j+}$, we know
$\lambda_{i,j}$ from (\ref{cd1}).
Substituting (\ref{cd1}) in (\ref{soft}), we can write
\begin{eqnarray}
\label{cd2}
\tilde{b}_{i,j} \, \Vert{\bf h}_i\Vert^2 & = & \Vert{\bf y} - {\bf H}{\bf d}_i^{j+} - \lambda_{i,j} {\bf h}_i\Vert^2 - \Vert{\bf y} - {\bf H}{\bf d}_i^{j+}\Vert^2  \nonumber \\
& = & \lambda_{i,j}^{^2} \Vert{\bf h}_i\Vert^2 - 2 \lambda_{i,j} {\bf h}_i^T ({\bf y} - {\bf H}{\bf d}_i^{j+}) \label{plus} \\
& = & -\lambda_{i,j}^{^2} \Vert{\bf h}_i\Vert^2 - 2 \lambda_{i,j} {\bf h}_i^T ({\bf y} - {\bf H}{\bf d}_i^{j-}) \label{minus}.
\end{eqnarray}
If $b_{i,j}=1$, then ${\bf d}_i^{j+}={\bf d}$ and substituting this in
(\ref{plus}) and dividing by $\Vert{\bf h}_i\Vert^2$, we get
\begin{eqnarray}
\label{cd3}
\tilde{b}_{i,j} & = &  \lambda_{i,j}^{^2} - 2 \lambda_{i,j} \frac{z_i}{({\bf G})_{i,i}}.
\end{eqnarray}
If $b_{i,j} = -1$, then ${\bf d}_i^{j-}={\bf d}$ and
substituting this in (\ref{minus}) and dividing by $\Vert{\bf h}_i\Vert^2$,
we get
\begin{eqnarray}
\label{cd4}
\tilde{b}_{i,j} & = & -\lambda_{i,j}^{^2} - 2 \lambda_{i,j} \frac{z_i}{({\bf G})_{i,i}}.
\end{eqnarray}
It is noted that ${\bf z}$ and {\bf G} are already available upon the
termination of the $M$-LAS algorithm, and hence the complexity of computing
$\tilde{b}_{i,j}$ in (\ref{cd3}) and (\ref{cd4}) is constant. Hence, the
overall complexity in computing the soft values for all the bits is
{\small $O(N_t\log_2 {\mathcal M})$}. We also see from (\ref{cd3}) and
(\ref{cd4})
that the magnitude of $\tilde{b}_{i,j}$ depends upon $\lambda_{i,j}$.
For large-size signal sets, the possible values of $\lambda_{i,j}$ will
also be large in magnitude. We therefore have to normalize $\tilde{b}_{i,j}$
for the turbo decoder to function properly. It has been observed through
simulations that normalizing $\tilde{b}_{i,j}$ by
$\big(\frac{\lambda_{i,j}}{2}\big)^2$ resulted in good performance.
In \cite{isit08}, we have shown that this soft decision output generation
method, when used in large V-BLAST systems, offers about 1 to 1.5 dB
improvement in coded BER performance compared to that achieved using
hard decision outputs from the $M$-LAS algorithm. We have observed similar
improvements in STBC MIMO systems also. In all coded BER simulations in
this paper, we use the soft outputs proposed here as inputs to the
decoder.

\section{BER Performance with Perfect CSIR}
\label{sec4}
In this section, we present the uncoded/turbo coded BER performance of
the proposed $M$-LAS detector in decoding non-orthogonal STBCs from CDA,
assuming perfect knowledge of CSI at the receiver\footnote{We will relax
this perfect channel knowledge assumption in the next section, where we
present an iterative detection/channel estimation scheme for the considered
large STBC MIMO system.}. In all the BER simulations in this section, we
have assumed that the fade remains constant over one STBC matrix duration
and varies i.i.d. from one STBC matrix duration to the other. We consider
two STBC designs; $i)$ `FD-ILL'
STBCs where $\delta=e^{\sqrt{5}\,{\bf j}}$, $t=e^{{\bf j}}$ in (11.a),
and $ii)$ `ILL-only' STBCs where $\delta=t=1$. The SNRs in all the BER
performance figures are the average received SNR per received antenna,
$\gamma$, defined in Sec. \ref{sec2} \cite{jafarkhani}. We have used
MMSE filter as the initial filter in all the simulations.

\subsection{Uncoded BER as a Function of Increasing $N_t=N_r$}
\label{sec41}
\begin{figure}
\hspace{-6mm}
\includegraphics[width=3.75in]{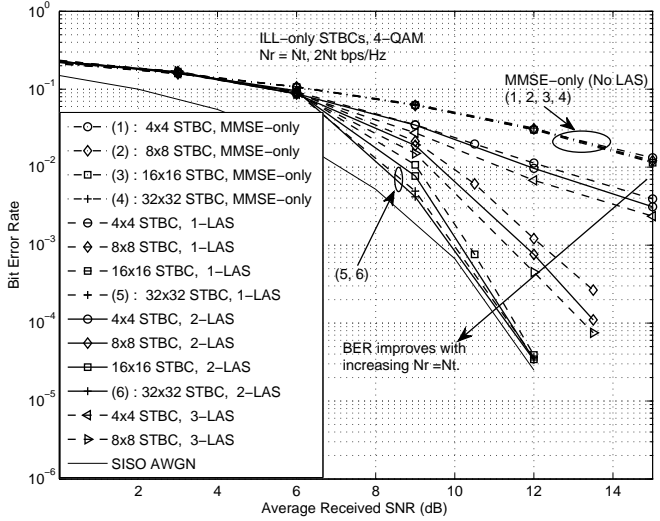}
\caption{
Uncoded BER of the proposed 1-LAS, 2-LAS and 3-LAS detectors for
{\bf ILL-only} STBCs for different $N_t=N_r$. 4-QAM, $2N_t$ bps/Hz.
{\em BER improves as $N_t=N_r$ increases and approaches SISO AWGN
performance for large $N_t=N_r$}.
}
\vspace{-2mm}
\label{fig2}
\end{figure}

In Fig. \ref{fig2}, we plot the uncoded BER performance of the proposed
1-, 2-, and 3-LAS algorithms in decoding ILL-only STBCs ($4\times 4$,
$8\times 8$, $16\times 16$, $32\times 32$ STBCs) for $N_t=N_r=4,8,16,32$
and 4-QAM. SISO AWGN performance (without fading) and MMSE-only
performance (i.e., without the search using LAS) are also plotted for
comparison. It can be seen that MMSE-only performance does not improve
with increasing STBC size (i.e., increasing $N_t=N_r$). However, it is
interesting to see that, when the proposed search using LAS is performed
following the MMSE operation, the performance improves for increasing
$N_t=N_r$, illustrating the performance benefit due to the proposed
search strategy. For example, though the LAS detector performs far from
SISO AWGN performance for small number of dimensions (e.g., $4\times 4,
8\times 8$ STBCs with 32 and 128 real dimensions, respectively), its large
system behavior at increased number of dimensions (e.g., $16\times 16$
and $32\times 32$ STBCs with 512 and 2048 real dimensions, respectively)
effectively renders near SISO AWGN performance; e.g., with $N_t=N_r=16,32$,
for BERs better than $10^{-3}$, the LAS detector performs very close to
SISO AWGN performance. We also observe that 3-LAS performs better than
2-LAS for $N_t=N_r=4,8$, and 2-LAS performs better than 1-LAS. Since close
to SISO AWGN performance is achieved with 1-, 2-, or 3-symbol update itself,
the cases of more than 3-symbol update, which will result in increased
complexity with diminishing returns in performance gain, are not considered
in the performance evaluation.

\begin{figure}
\hspace{-6mm}
\includegraphics[width=3.75in]{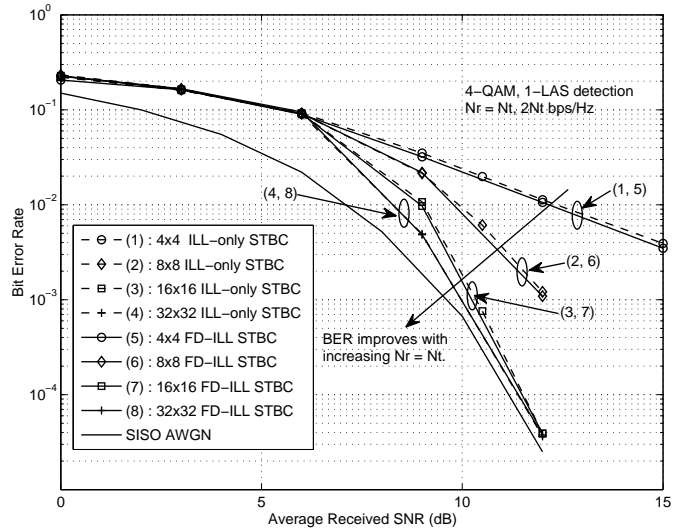}
\caption{
Uncoded BER comparison between {\bf FD-ILL} and {\bf ILL-only}
STBCs for different $N_t=N_r$. 4-QAM, $2N_t$ bps/Hz, 1-LAS
detection. {\em ILL-only STBCs perform almost same as FD-ILL STBCs.}
}
\vspace{-2mm}
\label{fig3}
\end{figure}

\subsection{Performance of FD-ILL Versus ILL-only STBCs}
In Fig. \ref{fig3}, we present uncoded BER performance comparison between
FD-ILL versus ILL-only STBCs for 4-QAM at different $N_t=N_r$ using 1-LAS
detection. The BER plots in Fig. \ref{fig3} illustrate that the performance
of ILL-only STBCs with 1-LAS detection for $N_t=N_r=4,8,16,32$ and 4-QAM
are almost as good as those of the corresponding FD-ILL STBCs. A similar
closeness between the performance of ILL-only and FD-ILL STBCs is observed
in the turbo coded BER performance as well, which is shown in Fig.
\ref{fig8} for a $16\times 16$ STBC with 4-QAM and turbo code rates of
1/3, 1/2 and 3/4. This is an interesting observation, since this suggests
that, in such cases, the computational complexity advantage with
$\delta=t=1$ in ILL-only STBCs can be taken advantage of without
incurring much performance loss compared to FD-ILL STBCs.

\subsection{Decoding and BER of Perfect Codes of Large Dimensions}
\begin{figure}
\hspace{-6mm}
\includegraphics[width=3.95in]{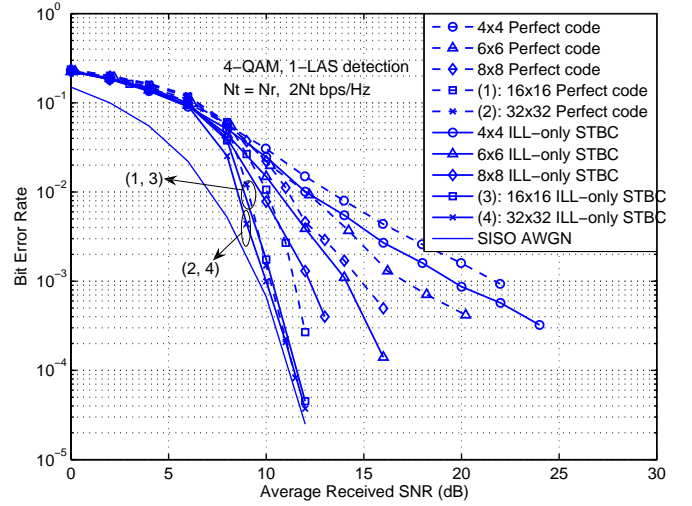}
\caption{
Uncoded BER comparison between {\bf perfect codes} and {\bf ILL-only}
STBCs for different $N_t=N_r$, 4-QAM, $2N_t$ bps/Hz, 1-LAS detection.
{\em For small
dimensions (e.g., $4\times 4$, $6\times 6$, $8\times 8$), perfect codes
with 1-LAS detection perform worse than ILL-only STBCs. For large
dimensions (e.g., $16\times 16$, $32\times 32$), ILL-only STBCs and
perfect codes perform almost same.}
}
\vspace{-2mm}
\label{fig4}
\end{figure}

\begin{figure}
\hspace{-6mm}
\includegraphics[width=3.75in]{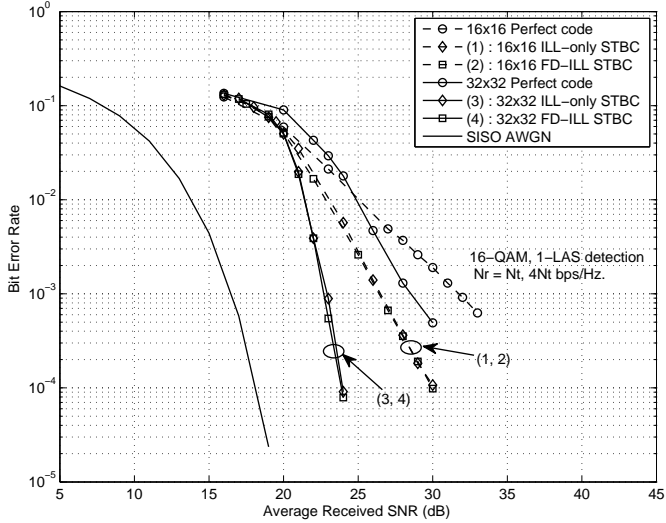}
\caption{
Uncoded BER comparison between {\bf perfect codes}, {\bf ILL-only}, and
{\bf FD-ILL} STBCs for $N_t=N_r=16,32$, {\bf 16-QAM}, $4N_t$ bps/Hz, 1-LAS
detection.  {\em For larger modulation alphabet sizes (e.g., 16 QAM),
perfect codes with 1-LAS detection perform poorer than ILL-only and FD-ILL
STBCs}.
}
\vspace{-2mm}
\label{fig5}
\end{figure}

While the STBC design in (11.a) offers both ILL and FD, {\em perfect
codes}\footnote{We note that the definition of perfect codes differ in
\cite{perf06} and \cite{perf07}. The perfect codes covered by the
definition in \cite{perf07} includes the perfect codes of \cite{perf06}
as a proper subclass. However, for our purpose of illustrating the
performance of the proposed detector in large STBC MIMO systems,
we refer to the codes in \cite{perf06} as well as  \cite{perf07}
as perfect codes.} under ML decoding
can provide coding gain in addition to ILL and FD \cite{gold05}-\cite{cda}.
Decoding of perfect codes has been reported in the literature for only up
to 5 antennas using sphere/lattice decoding \cite{perf07}. The complexity
of these decoders are prohibitive for decoding large-sized perfect codes,
although large-sized codes are of interest from a high spectral efficiency
view point. We note that, because of its low-complexity attribute, the
proposed $M$-LAS detector is able to decode perfect codes of large dimensions.
In Figs. \ref{fig4} and \ref{fig5}, we present the simulated BER performance
of perfect codes in comparison with those of ILL-only and FD-ILL STBCs for
up to 32 transmit antennas using 1-LAS detector.

In Fig. \ref{fig4}, we show uncoded BER comparison between perfect
codes and ILL-only STBCs for different $N_t=N_r$ and 4-QAM using 1-LAS
detection. The $4\times 4$ and $6\times 6$ perfect codes are from
\cite{perf06}, and the $8\times 8$, $16\times 16$ and $32\times 32$
perfect codes are from \cite{perf07}. From Fig. \ref{fig4}, it can be seen
that the 1-LAS detector achieves better performance for ILL-only STBCs than
for perfect codes, when codes with small number of transmit antennas are
considered (e.g., $N_t=4,6,8$). While perfect codes are expected to perform
better than ILL-only codes under ML detection for any $N_t$, we observe the
opposite behavior under 1-LAS detection for small $N_t$ (i.e., ILL-only
STBCs performing better than perfect codes for small
dimensions). This behavior could be attributed to the nature of the LAS
detector, which achieves near-optimal performance only when the number
of dimensions is large\footnote{In \cite{proof}, we have presented an
analytical proof that the bit error performance of 1-LAS detector for
V-BLAST with 4-QAM in i.i.d. Rayleigh fading converges to that of the
ML detector as $N_t,N_r \rightarrow \infty$, keeping $N_t=N_r$.}, and it
appears that, in the detection process, LAS is more effective in
disentangling the symbols in STBCs when $\delta=t=1$ (i.e., in ILL-only
STBCs) than in perfect codes. The performance gap between perfect codes
and ILL-only STBCs with 1-LAS detection diminishes for increasing
code sizes such that the performance for $32\times 32$ perfect code and
ILL-only STBC with 4-QAM are almost same and close to the SISO AWGN
performance. In Fig. \ref{fig5}, we show a similar comparison between
perfect codes, ILL-only and FD-ILL only STBCs when larger modulation
alphabet sizes (e.g., 16-QAM) are used in the case of $16\times 16$ and
$32\times 32$ codes. It can be seen that with higher-order QAM like
16-QAM, perfect codes with 1-LAS detection perform poorer than ILL-only
and FD-ILL STBCs, and that ILL-only and FD-ILL STBCs perform almost same
and close to the SISO AWGN performance. The results in Figs. \ref{fig4}
and \ref{fig5} suggest that, with 1-LAS detection, owing to the
complexity advantage and good performance in using $\delta=t=1$, ILL-only
STBCs can be a good choice for practical large STBC MIMO systems
\cite{hassibi2},\cite{cpa}.

\subsection{Comparison with Other Large-MIMO Architecture/ \hspace{-1.0mm}Detector Combinations}
\label{sec44}
\begin{figure}
\hspace{-4mm}
\includegraphics[width=3.6in]{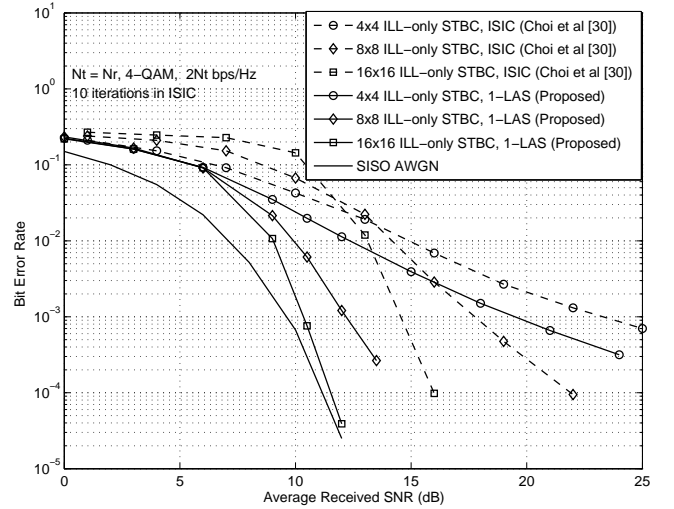}
\caption{
Uncoded BER comparison between the proposed 1-LAS algorithm
and the ISIC algorithm in \cite{choi} for {\bf ILL-only} STBCs for
different $N_t=N_r$. 4-QAM, $2N_t$ bps/Hz. MMSE initial vectors for
both 1-LAS and ISIC. {\em 1-LAS performs significantly better than ISIC
in \cite{choi}.}
}
\vspace{-2mm}
\label{fig6}
\end{figure}

In \cite{choi}, Choi {\em et al} have presented an iterative soft
interference cancellation (ISIC) scheme for multiple antenna systems,
derived based on maximum a posteriori (MAP) criterion. We compared the
performance of the ISIC scheme in \cite{choi} with that of the proposed
1-LAS algorithm in detecting $4\times 4$, $8\times 8$ and $16\times 16$
ILL-only STBCs with $N_t=N_r$ and 4-QAM. Figure \ref{fig6} shows this
performance comparison. In \cite{choi}, zero-forcing vector was used
as the initial vector in the ISIC scheme. However, performance is better
with MMSE initial vector. Since we used MMSE initial vector for $1$-LAS,
we have used MMSE initial vector for the ISIC algorithm as well. Also,
in \cite{choi}, 4 to 5 iterations were shown to be good enough for the
ISIC algorithm to converge. In our simulations of the ISIC algorithm,
we used 10 iterations. Two key observations can be made from Fig.
\ref{fig6}: $i)$ like the $1$-LAS algorithm, the ISIC algorithm also
shows large system behavior (i.e., improved BER for increasing $N_t=N_r$),
and 2) the proposed 1-LAS algorithm outperforms the ISIC algorithm by
about 3 to 5 dB at $10^{-3}$ uncoded BER. In addition, the complexity
of the ISIC scheme is higher than the proposed scheme (see the
complexity comparison in Table I).

\begin{figure}
\hspace{-6mm}
\includegraphics[width=3.80in]{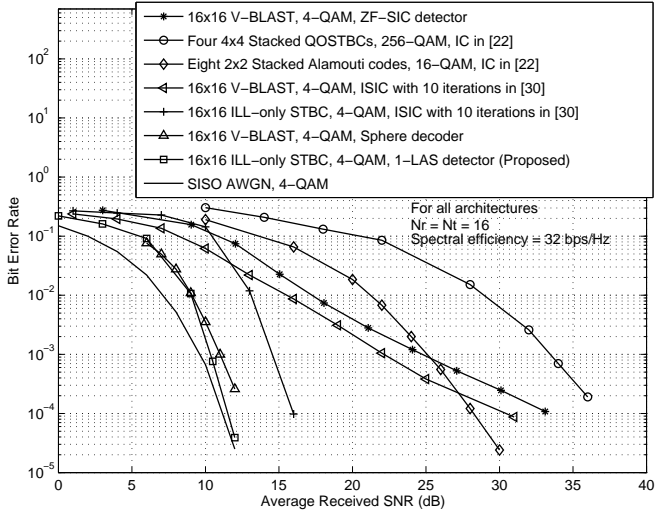}
\caption{
Uncoded BER comparison between different large-MIMO
architecture/detector combinations for given number of transmit/receive
antennas ($N_t=N_r=16$) and spectral efficiency (32 bps/Hz). {\em Proposed
scheme performs better than other architecture/detector
combinations considered. It outperforms them in complexity as well
(see Table I).}
}
\vspace{-2mm}
\label{fig7}
\end{figure}

Next, we compare the proposed large-MIMO architecture using STBCs from
CDA and $M$-LAS detection with other large-MIMO architectures and
associated detectors reported in the literature. Large-MIMO architectures
that use stacking of multiple small-sized STBCs and interference
cancellation (IC) detectors for these schemes have been investigated
in \cite{kazem},\cite{prasad},\cite{tan}. Here, we compare different
architecture/detector combinations, fixing the total number of
transmit/receive antennas and spectral efficiency to be same in all
the considered combinations. Specifically, we fix $N_t=N_r=16$ and
a spectral efficiency of 32 bps/Hz for all the combinations.
We compare the
following seven different architecture/detector combinations which
use the same $N_t=N_r=16$ and achieve 32 bps/Hz spectral efficiency
(see Table I):
$i)$ proposed scheme using $16\times 16$ ILL-only STBC (rate-16) with
4-QAM and 1-LAS detection,
$ii)$ $16\times 16$ ILL-only STBC (rate-16) with 4-QAM and ISIC
algorithm in \cite{choi} with 10 iterations,
$iii)$ four $4\times 4$ stacked QOSTBCs (rate-1) with 256-QAM and
IC algorithm presented in \cite{kazem},
$iv)$ eight $2\times 2$ stacked Alamouti codes (rate-1) with 16-QAM
and IC algorithm in \cite{kazem},
$v)$ $16\times 16$ V-BLAST scheme (rate-16) with 4-QAM and sphere
decoder (SD),
$vi)$ $16\times 16$ V-BLAST scheme (rate-16) with 4-QAM and ZF-SIC
detector, and
$vii)$ $16\times 16$ V-BLAST scheme (rate-16) with 4-QAM and ISIC
algorithm in \cite{choi}.
We present the BER performance comparison of these different
combinations in Fig. \ref{fig7}. We also obtained the complexity
numbers (in number of real operations per bit) from simulations
for these different combinations at an uncoded BER of $5\times 10^{-2}$;
these numbers are presented in Table I, along with the SNRs at which
$5\times 10^{-2}$ uncoded BER is achieved. The following interesting
observations can be made from Fig. \ref{fig7} and Table I:

\begin{table*}[t]
\begin{center}
\begin{tabular}{|c|c|c|c|}
\hline
&               & Complexity & SNR required      \\
No.&  Large-MIMO Architecture/Detector Combinations & (in \# real operations  & to achieve $5\times 10^{-2}$ \\
& (fixed $N_t=N_r=16$ and 32 bps/Hz & per bit) at $5\times 10^{-2}$ & uncoded BER  \\
& for all combinations) & uncoded BER & (from Fig. \ref{fig7}) \\ \hline \hline
& {\bf ${\bf 16\times 16}$ ILL-only CDA STBC (rate-16)}, &              & \\
$i)$ & {\bf 4-QAM and 1-LAS detection}            & ${\bf 3.473 \times 10^{3}}$ & {\bf 6.8 dB} \\
& {\bf [Proposed scheme] }                &             & \\ \hline
$ii)$ & $16\times 16$ ILL-only CDA STBC (rate-16), &            & \\
& 4-QAM and ISIC algorithm in \cite{choi}   & $1.187 \times 10^{5}$ & 11.3 dB \\ \hline
$iii)$ & Four $4\times 4$ stacked rate-1 QOSTBCs,   & & \\
& 256-QAM and IC algorithm in \cite{kazem}  & $5.54\times 10^6$ & 24 dB \\ \hline
$iv)$ & Eight $2\times 2$ stacked rate-1 Alamouti codes, & & \\
& 16-QAM and IC algorithm in \cite{kazem}   & $8.719\times 10^{3}$ & 17 dB \\ \hline
$v)$ & $16\times 16$ V-BLAST (rate-16) scheme,    & & \\
& 4-QAM and sphere decoding               & $4.66\times 10^{4}$ & 7 dB \\ \hline
$vi)$ & $16\times 16$ V-BLAST (rate-16) scheme,           & & \\
& 4-QAM and V-BLAST detector (ZF-SIC)     & $1.75\times 10^{4}$ & 13 dB \\ \hline
$vii)$ & $16\times 16$ V-BLAST (rate-16) scheme,   & & \\
& 4-QAM and ISIC algorithm in \cite{choi}   & $7.883\times 10^{3}$ & 10.6 dB \\ \hline
\end{tabular}
\label{tab2}
\vspace{2mm}
\caption{
Complexity and performance comparison of different large-MIMO
architecture/detector combinations, all with $N_t=N_r=16$ and achieving 
32 bps/Hz spectral efficiency. {\em Proposed scheme outperforms the other 
considered architectures/detectors both in terms of performance
as well as complexity}. }
\end{center}
\label{table1}
\end{table*}

\begin{itemize}
\item
the proposed scheme \big(combination $i)$\big) significantly outperforms
the stacked architecture/IC detector combinations presented in \cite{kazem}
\big(combinations $iii)$ and $iv)$\big); e.g.,
at $5\times 10^{-2}$ uncoded BER, the proposed scheme performs better
than the stacked architecture/IC in \cite{kazem} by 17 dB (for four
$4\times 4$ QOSTBCs) and 10 dB (for eight $2\times 2$ Alamouti codes).
Also, the proposed scheme achieves this significant performance advantage at
a much lesser complexity than those of the stacked architecture/IC 
combinations (see Table I).
\item
the proposed scheme performs slightly better than the V-BLAST/sphere
decoder combination \big(combination $v)$\big); 6.8 dB in proposed scheme
versus 7 dB in V-BLAST with sphere decoding at $5\times 10^{-2}$ uncoded
BER. Importantly, the
proposed scheme enjoys a significant complexity advantage (by more than
an order) over the V-BLAST/sphere decoder combination.
\item
the ISIC algorithm in \cite{choi} applied to ILL-only STBC detection
(combination $ii)$) is inferior to the proposed scheme in both performance
(by about 4.5 dB at $5\times 10^{-2}$ uncoded BER) as well as complexity
(by about two orders).
\item
the ISIC algorithm in \cite{choi} applied to $16\times 16$ V-BLAST
detection \big(combination $vii)$\big) is also inferior to the proposed
scheme in BER performance (by about 3.8 dB at $5\times 10^{-2}$ uncoded
BER) as well as complexity (by about a factor of 2).
\item
comparing the stacked architecture/IC combinations with V-BLAST/ZF-SIC
\big(combination $vi)$\big)
and V-BLAST/ISIC combinations, we see that although the
diversity orders achieved in stacked architecture/IC combinations
are high (see their slopes at high SNRs in Fig. \ref{fig7}), V-BLAST
with ZF-SIC and ISIC detectors perform much better at low and medium SNRs.
\end{itemize}
In summary, the proposed scheme outperforms the other considered
architecture/detector combinations both in terms of performance 
as well as complexity.

\subsection{Turbo Coded BER and Nearness-to-Capacity Results}
\begin{figure}
\hspace{-6mm}
\includegraphics[width=3.70in]{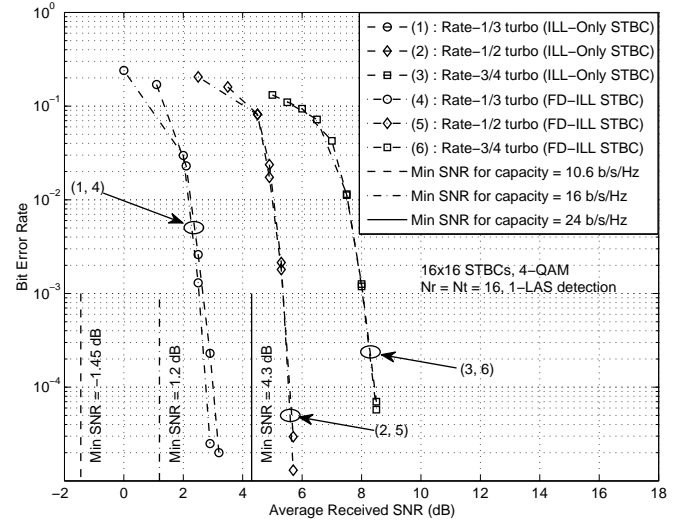}
\caption{
Turbo coded BER of 1-LAS detector for $16\times 16$
{\bf FD-ILL} and {\bf ILL-only} STBCs. $N_t=N_r=16$, 4-QAM,
turbo code rates: 1/3, 1/2, 3/4 {\bf(10.6, 16, 24 bps/Hz)}. {\em 1-LAS
detector performs close to within 4 dB from capacity}. {\em ILL-only
STBCs preform as good as FD-ILL STBCs.}
}
\vspace{-2mm}
\label{fig8}
\end{figure}

Next, we evaluated the turbo coded BER performance of the proposed scheme.
In all the coded BER simulations, we fed the soft outputs presented in
Sec. \ref{sec_soft} as input to the turbo decoder. In Fig. \ref{fig8},
we plot the turbo coded BER of the 1-LAS detector in decoding $16\times 16$
FD-ILL and ILL-only STBCs, with $N_t=N_r=16$, 4-QAM and turbo code rates 1/3
(10.6 bps/Hz), 1/2 (16 bps/Hz), 3/4 (24 bps/Hz). The minimum SNRs required
to achieve these capacities in a $16\times 16$ MIMO channel (obtained by
evaluating the ergodic capacity expression in \cite{tela99} through
simulation) are also shown. It can be seen that the 1-LAS detector
performs close to within just about 4 dB from capacity, which is very
good in terms of nearness-to-capacity considering the high spectral
efficiencies achieved. It can also be seen that the coded BER performance
of FD-ILL and ILL-only STBCs are almost the same for the system parameters
considered.

\subsection{Effect of MIMO Spatial Correlation}
In generating the BER results in Figs. \ref{fig2} to \ref{fig8}, we have
assumed i.i.d. fading. However, MIMO propagation conditions witnessed in
practice often render the i.i.d. fading model as inadequate. More realistic
MIMO channel models that take into account the scattering environment,
spatial correlation, etc., have been investigated in the literature
\cite{mimo3},\cite{mimo1}. For example, spatial correlation at the
transmit and/or receive side can affect the rank structure of the MIMO
channel resulting in degraded MIMO capacity \cite{mimo1}.
The structure of scattering in the propagation environment can also affect
the capacity \cite{mimo3}. Hence, it is of interest to investigate the
performance of the $M$-LAS detector in more realistic MIMO channel models. To
this end, we use the non-line-of-sight (NLOS) correlated MIMO channel model
proposed by Gesbert {\em et al}\,\footnote{Please see \cite{mimo3} for more 
elaborate details of the spatially correlated MIMO channel model. We note 
that this model can be appropriate in application scenarios like high data 
rate wireless IPTV/HDTV distribution using high spectral efficiency 
large-MIMO links, where large $N_t$ and $N_r$ can be placed at the base 
station (BS) and customer premises equipment (CPE), respectively.} in 
\cite{mimo3}, and evaluate the effect of spatial correlation on the BER 
performance of the $M$-LAS detector \cite{pimrc08}.

\begin{figure}
\hspace{-0mm}
\includegraphics[width=3.40in]{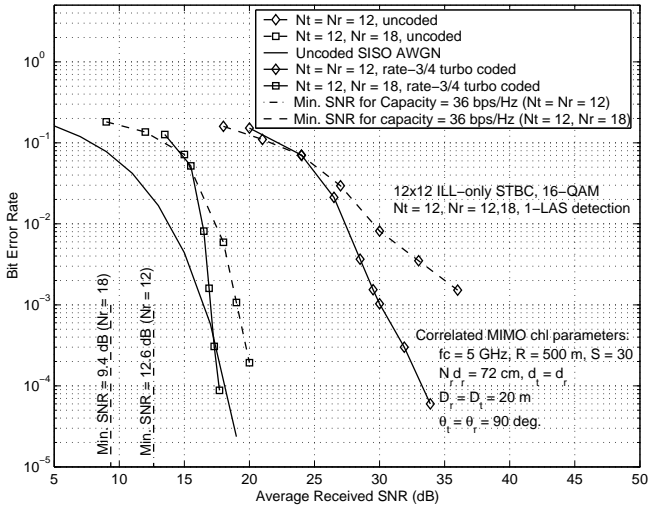}
\caption{
Effect of $N_r>N_t$ in correlated MIMO
fading in \cite{mimo3} keeping $N_rd_r$ constant and $d_t=d_r$. $N_rd_r=72$ cm,
$f_c=5$ GHz, $R=500$ m, $S=30$, $D_t=D_r=20$ m, $\theta_t=\theta_r=90^\circ$,
$12\times 12$ {\bf ILL-only} STBC, $N_t=12$, $N_r=12,18$, {\bf 16-QAM},
rate-3/4 turbo code, {\bf 36 bps/Hz}. {\em Increasing \# receive dimensions
alleviates the loss due to spatial correlation}.
}
\vspace{-2mm}
\label{fig9}
\end{figure}

We consider the following parameters\footnote{The parameters used in the
model in \cite{mimo3} include: $N_t, N_r:$ \# transmit and receive
(omni-directional) antennas; $d_t,d_r$: spacing between antenna elements
at the transmit side and at the receive side; $R$: distance between
transmitter and receiver, $D_t,D_r$: transmit and receive scattering radii;
$S$: number of scatterers on each side; $\theta_t,\theta_r$: angular spread
at the transmit and receiver sides, and $f_c,\lambda$: carrier
frequency, wavelength.} in the simulations: $f_c=5$ GHz, $R=500$ m, $S=30$,
$D_t=D_r=20$ m, $\theta_t=\theta_r=90^\circ$, and $d_t=d_r=2\lambda/3$.
For $f_c=5$ GHz, $\lambda=6$ cm and $d_t=d_r=4$ cm. In Fig. \ref{fig7},
we plot the BER performance of the 1-LAS detector in decoding $16 \times 16$
ILL-only STBC with $N_t=N_r=16$ and 16-QAM. Uncoded BER as well as rate-3/4
turbo coded BER (48 bps/Hz spectral efficiency) for i.i.d. fading as well
as correlated fading are shown. In addition, from the MIMO capacity formula
in \cite{tela99}, we evaluated the theoretical minimum SNRs required to
achieve a capacity of 48 bps/Hz in i.i.d. as well as correlated fading,
and plotted them also in Fig. \ref{fig7}. It is seen that the minimum SNR
required to achieve a certain capacity (48 bps/Hz) gets increased for
correlated fading compared to i.i.d. fading. From the BER plots in Fig.
\ref{fig7}, it can be observed that at an uncoded BER of $10^{-3}$, the
performance in correlated fading degrades by about 7 dB compared that in
i.i.d. fading. Likewise, at a rate-3/4 turbo coded BER of $10^{-4}$, a
performance loss of about 6 dB is observed in correlated fading compared
to that in i.i.d. fading. In terms of nearness to capacity, the vertical
fall of the coded BER for i.i.d. fading occurs at about 24 dB SNR, which
is about 13 dB away from theoretical minimum required SNR of 11.1 dB.
With correlated fading, the detector is observed to perform close to
capacity within about 18.5 dB. One way to alleviate such degradation in
performance due to spatial correlation can be by providing more number
of dimensions at the receive side, which is highlighted in Fig. \ref{fig9}.

Figure \ref{fig9} illustrates that the 1-LAS detector can achieve substantial
improvement in uncoded as well as coded BER performance in decoding
$12\times 12$ ILL-only STBC by increasing $N_r$ beyond $N_t$
for 16-QAM in correlated fading. In the simulations, we have maintained
$N_rd_r=72$ cm and $d_t=d_r$ in both the cases of symmetry (i.e., {\small
$N_t=N_r=12$}) as well as asymmetry (i.e., {\small $N_t=12,N_r=18$}).
By comparing the 1-LAS detector performance with {\small [$N_t=N_r=12$]}
versus {\small [$N_t=12,N_r=18$]}, we observe that the uncoded BER
performance with {\small [$N_t=12,N_r=18$]} improves by about 17 dB
compared to that of {\small [$N_t=N_r=12$]} at $2\times 10^{-3}$ BER. Even
the uncoded BER performance with {\small [$N_t=12,N_r=18$]} is significantly
better than the coded BER performance with {\small [$N_t=N_r=12$]} by about
11.5 dB at $10^{-3}$ BER. This improvement is essentially due to the ability
of the 1-LAS detector to effectively pick up the additional diversity orders
provided by the increased number of receive antennas. With a rate-3/4 turbo
code (i.e., 36 bps/Hz), at a coded BER of $10^{-4}$, the 1-LAS
detector achieves a significant performance improvement of about 13 dB with
[$N_t=12,N_r=18$] compared to that with {\small [$N_t=N_r=12$]}. With
{\small [$N_t=12,N_r=18$]}, the vertical fall of coded BER is such that
it is only about 8 dB from the theoretical minimum SNR needed to achieve
capacity. This points to the potential for realizing high spectral efficiency
multi-gigabit large-MIMO systems that can achieve good performance even in
the presence of spatial correlation.
We further remark that transmit correlation in MIMO fading can be exploited
by using non-isotropic inputs (precoding) based on the knowledge of the
channel correlation matrices \cite{mimo4}-\cite{mimo6}. While
\cite{mimo4}-\cite{mimo6} propose precoders in conjunction with
orthogonal/quasi-orthogonal small MIMO systems in correlated Rayleigh/Ricean
fading, design of precoders for large-MIMO systems can be investigated as 
future work.

\section{Iterative Detection/Channel Estimation}
\label{sec5}
In this section, we relax the perfect CSIR assumption made in the previous
section, and estimate the channel matrix based on a training-based iterative
detection/channel estimation scheme \cite{icc09}.
Training-based schemes, where a pilot signal known to the transmitter
and the receiver is sent to get a rough estimate of the channel (training
phase) has been studied for STBC MIMO systems in \cite{est1}-\cite{est4}.
Here, we adopt a training-based approach for channel estimation in
large STBC MIMO systems. In the considered training-based channel
estimation scheme, transmission
is carried out in frames, where one $N_t\times N_t$ pilot matrix,
${\bf X}_c^{(\text{P})} \in {\mathbb C}^{N_t\times N_t}$, for training
purposes, followed by $N_d$ data STBC matrices,
${\bf X}_c^{(i)}\in \mathbb C^{N_t\times N_t},\ i=1,2,...,N_d$,
are sent in each frame as shown in Fig. \ref{fig11}. One frame length, $T$,
(taken to be the channel coherence time) is $T=(N_d+1)N_t$ channel uses.
A frame of transmitted pilot and data matrices is of dimension
$N_t\times N_t(1+N_d)$, which can be written as
\begin{eqnarray}
{\bf {\cal X}}_c & = & \left[{\bf X}_c^{\text{(P)}}\  {\bf X}_c^{(1)}\  {\bf X}_c^{(2)}\  {\bf \cdots}\  {\bf X}_c^{(N_d)}\right].
\label{calx}
\end{eqnarray}
As in \cite{hh03}, let $\gamma_p$ and $\gamma_d$ denote the average SNR
during pilot and data phases, respectively, which are related to the
average received SNR $\gamma$ as $\gamma(N_d+1)=\gamma_p+N_d\gamma_d$.
Define $\beta_p \Define \frac{\gamma_p}{\gamma}$, and
$\beta_d \Define \frac{\gamma_d}{\gamma}$. Let $E_s$ denote the average
energy of the transmitted symbol during the data phase. The average
received signal power during the data phase is given by
${\mathbb E}\big[ \mbox{tr}\big({\bf X}_c^{(\text{i})}{{\bf X}_c^{(\text{i})}}^H\big)\big]=N_t^2E_s$,
and the average received signal power during the pilot phase is
${\mathbb E}\big[\mbox{tr}\big({\bf X}_c^{(\text{P})}{{\bf X}_c^{(\text{P})}}^H\big)\big]=\frac{N_t^2E_s\beta_p}{\beta_d} = \mu N_t$,
where $\mu \Define \frac{N_tE_s \beta_p}{\beta_d}$.
For optimal training, the pilot matrix should be such that
${\bf X}_c^{\text{(P)}} {{\bf X}_c^{\text{(P)}}}^H = \mu {\bf I}_{N_t}$
\cite{hh03}. As in Sec. \ref{sec2}, let
${\bf H}_c \in {\mathbb C}^{N_r\times N_t}$ denote the channel matrix, which
we want to estimate. We assume block fading, where the channel gains remain
constant over one frame consisting of $(1+N_d)N_t$ channel uses, which can
be viewed as the channel coherence time. This assumption can be valid in
slow fading fixed wireless applications (e.g., as in possible applications
like BS-to-BS backbone connectivity and BS-to-CPE wireless IPTV/HDTV
distribution).
For this training-based system and channel model, Hassibi and Hochwald
presented a lower bound on the capacity in \cite{hh03}; we will illustrate
the nearness of the performance achieved by the proposed iterative
detection/estimation scheme to this bound. The received frame is of
dimension $N_r\times N_t(1+N_d)$, and can be written as
\begin{eqnarray}
\label{gen_model_eqn}
\hspace{-6mm}
{\bf {\cal Y}}_c &\hspace{-1mm} = & \hspace{-1mm} \left[{\bf Y}_c^{\text{(P)}}\
{\bf Y}_c^{(1)}\
{\bf Y}_c^{(2)}\
{\bf \cdots }\
{\bf Y}_c^{(N_d)}\
\right] \,\, = \,\, {\bf H}_c\, {\bf {\cal X}}_c + {\bf {\cal N}}_c\ ,
\end{eqnarray}
where
{\small
${\bf {\cal N}}_c = \left[{\bf N}_c^{\text{(P)}}\
{\bf N}_c^{(1)}\
{\bf N}_c^{(2)}\
{\bf \cdots }\
{\bf N}_c^{(N_d)}\
\right]$
}
is the $N_r\times N_t(1+N_d)$ noise matrix and its entries are
modeled as i.i.d.
$\mathcal C \mathcal N(0,\sigma^2=\frac{N_tE_s}{\gamma \beta_d})$.
Equation (\ref{gen_model_eqn}) can be decomposed into two parts, namely,
the pilot matrix part and the data matrices part, as
\begin{eqnarray}
\label{decom_model1}
{\bf Y}_c^{\text{(P)}} & = & {\bf H}_c {\bf X}_c^{\text{(P)}} + {\bf N}_c^{\text{(P)}},
\end{eqnarray}
\begin{eqnarray}
\label{decom_model2}
\hspace{-6mm}
{\bf Y}_c^{\text{(D)}} & = & \left[{\bf Y}_c^{(1)}\  {\bf Y}_c^{(2)}\  {\bf \cdots}\  {\bf Y}_c^{(N_d)}\right] \nonumber \\
& \hspace{-20mm} = & \hspace{-12mm} {\bf H}_c \left[{\bf X}_c^{(1)}\  {\bf X}_c^{(2)}\ {\bf \cdots}\ {\bf X}_c^{(N_d)}\right] + \left[{\bf N}_c^{(1)}\  {\bf N}_c^{(2)}\  {\bf \cdots}\  {\bf N}_c^{(N_d)}\right]\hspace{-1mm}.
\end{eqnarray}

\begin{figure}
\hspace{-0mm}
\includegraphics[width=3.40in]{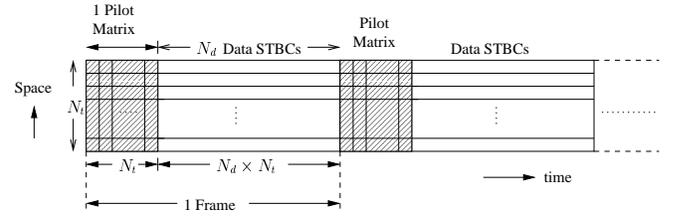}
\caption{
Transmission scheme with one pilot matrix followed by
$N_d$ data STBC matrices in each frame.
}
\vspace{-2mm}
\label{fig10}
\end{figure}

\subsection{MMSE Estimation Scheme}
\label{seca}
A straight-forward way to achieve detection of data symbols with
estimated channel coefficients is as follows:
\begin{enumerate}
\item   Estimate the channel gains via an {\em MMSE estimator} from the
        signal received during the first $N_t$ channel uses (i.e.,
        during pilot transmission); i.e., given ${\bf Y}_c^{\text{(P)}}$
        and ${\bf X}_c^{\text{(P)}}$, an estimate of the channel matrix
        ${\bf H}_c$ is found as
\begin{eqnarray}
\label{mmse1}
\hspace{-6mm}
{\bf H}_c^{est} & \hspace{-1.5mm} = & \hspace{-1.5mm} {\bf Y}_c^{\text{(P)}}\ ({\bf X}_c^{\text{(P)}})^H \left[\sigma^2 {\bf I}_{N_t} + {\bf X}_c^{\text{(P)}} ({\bf X}_c^{\text{(P)}})^H \right]^{-1}\hspace{-2mm}.
\end{eqnarray}
\item   Use the above ${\bf H}_c^{est}$ in place of ${\bf H}_c$ in the LAS
        algorithm (as described in Sections \ref{sec2} and \ref{sec3})
        and detect the transmitted data symbols.
\end{enumerate}
We refer to the above scheme as the {\em `MMSE estimation scheme.'}
In the absence of the knowledge of $\sigma^2$, a zero-forcing estimate
can be obtained at the cost of some performance loss compared to the
MMSE estimate. The performance of the estimator can be improved
by using a cyclic minimization technique for minimizing the ML metric
\cite{stoica}.

\subsection{Proposed Iterative Detection/Estimation Scheme}
\label{secb}
Techniques that employ iterations between channel estimation and detection
can offer improved performance. Iterative receiver algorithms are attractive
to achieve a good tradeoff between performance and complexity
\cite{lampe}-\cite{ralf1}. In \cite{lampe}-\cite{hu}, receivers
that iterate between channel estimation, multiuser detection and channel
decoding in coded CDMA systems are presented. Similar iterative techniques
in the context of MIMO and MIMO-OFDM systems are presented in
\cite{ralf0}-\cite{ralf1}. Here, we propose an iterative scheme, where
we iterate between channel estimation and detection in the considered
large STBC MIMO system. The proposed scheme works as follows:
\begin{enumerate}
\item   Obtain an initial estimate of the channel matrix using the MMSE
        estimator in (\ref{mmse1}) from the pilot part.
\item   Using the estimated channel matrix, detect the data STBC matrices
        ${\bf X}_c^{(i)}$, $i=1,2,\cdots,N_d$ using the LAS detector.
        Substituting these detected STBC matrices into (\ref{calx}),
        form ${\bf {\cal X}}_c^{est}$.
\item   Re-estimate the channel matrix using ${\bf {\cal X}}_c^{est}$ from
        the previous step, via
\begin{eqnarray}
\label{mmse2}
\hspace{-6mm}
{\bf H}_c^{est} & \hspace{-1.25mm} = & \hspace{-1.25mm} {\bf {\cal Y}}_c ({\bf {\cal X}}_c^{est})^H \left[\sigma^2 {\bf I}_{N_t} +  {\bf {\cal X}}_c^{est} ({\bf {\cal X}}_c^{est})^H  \right]^{-1}\hspace{-2mm}.
\end{eqnarray}
\item   Iterate steps 2 and 3 for a specified number of iterations.
\end{enumerate}
The total complexity of obtaining the MMSE estimate of the channel matrix
${\bf H}_c^{est}$ in (\ref{mmse1}) and (\ref{mmse2}) is
$O(N_t^2N_r)+O(N_t^3)$, which is less than the total complexity
of 1-LAS detection of $O(N_t^4\log N_t)$ for ILL-only STBCs.

\begin{figure}
\hspace{-6mm}
\includegraphics[width=3.90in]{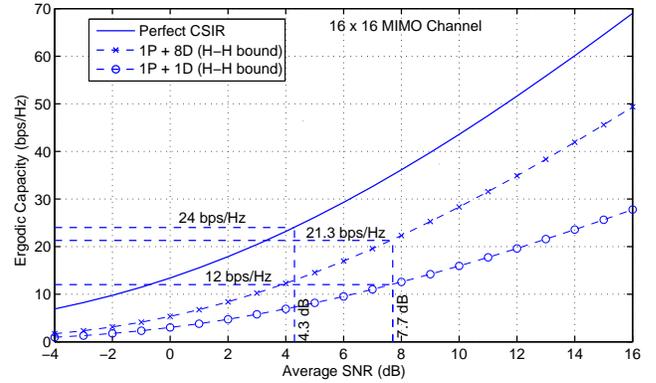}
\caption{
Hassibi-Hochwald (H-H) capacity bound for
1P+8D ($T=144,\tau=16,\beta_p=\beta_d=1$) and
1P+1D ($T=32,\tau=16,\beta_p=\beta_d=1$) training for a
$16 \times 16$ MIMO channel. Perfect CSIR capacity is also shown.
}
\vspace{-2mm}
\label{fig11}
\end{figure}

\subsection{BER Performance with Estimated CSIR}
We evaluated the BER performance of the 1-LAS detector using estimated
CSIR, where we estimate the channel gain matrix through the training-based
estimation schemes describ\-ed in the previous two subsections. We consider
the BER performance under three scenarios, namely, $i)$ under perfect
CSIR, $ii)$ under CSIR estimated using the MMSE estimation scheme in Sec.
\ref{seca}, and $iii)$ under CSIR estimated using the iterative
detection/estimation scheme in Sec. \ref{secb}. In the case of estimated
CSIR, we show plots for 1P+$N_d$D training, where by 1P+$N_d$D training
we mean a training scheme with a frame size of $1+N_d$ matrices, with
1 pilot matrix followed $N_d$ data STBC matrices from CDA. For this
1P+$N_d$D training scheme, a lower bound on the capacity is
given by \cite{hh03}

\vspace{-2mm}
{\footnotesize
\begin{eqnarray}
\hspace{-6mm}
C  & \hspace{-1.75mm} \geq & \hspace{-1.75mm} \frac{T-{\tau}}{T} \ \mathbb{E}\left[\mbox{logdet}\left({\bf I}_{N_t}+\frac{\gamma^2{\beta}_d {\beta}_p \tau}{N_t(1+\gamma{\beta}_d)+ \gamma{\beta}_p \tau} \
\frac{ \hat{{\bf H}}_c \hat{{\bf H}}_c^H }{ N_t \sigma_{{\hat{\bf H}}_c}^2 }
\right) \right]\hspace{-0.95mm},
\label{trg_bound}
\end{eqnarray}
}

\begin{figure}
\hspace{-6mm}
\includegraphics[width=4.00in]{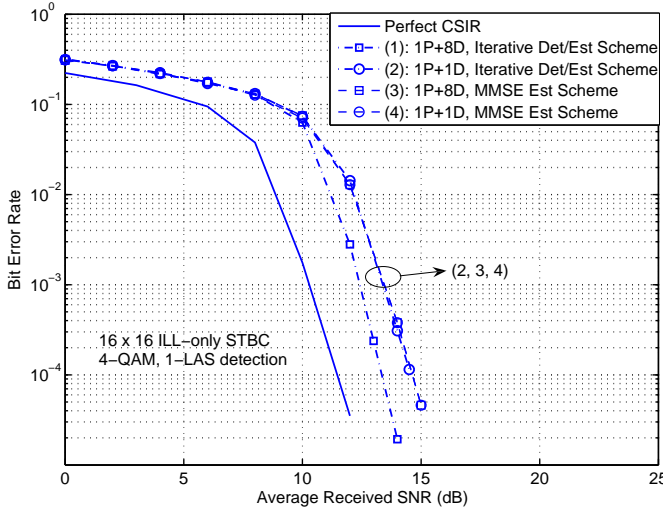}
\caption{
Uncoded BER of 1-LAS detector for $16\times 16$ {\bf ILL-only} STBC with
$i)$ perfect CSIR,
$ii)$ CSIR using MMSE estimation scheme, and $iii)$ CSIR using iterative
detection/channel estimation scheme (4 iterations). $N_t=N_r=16$, 4-QAM,
1P+1D \big($T=32,\tau=16,\beta_p=\beta_d=1$\big) and 1P+8D
\big($T=144,\tau=16,\beta_p=\beta_d=1$\big) training.
}
\vspace{-2mm}
\label{fig12}
\end{figure}

\begin{figure}
\hspace{-6mm}
\includegraphics[width=4.00in]{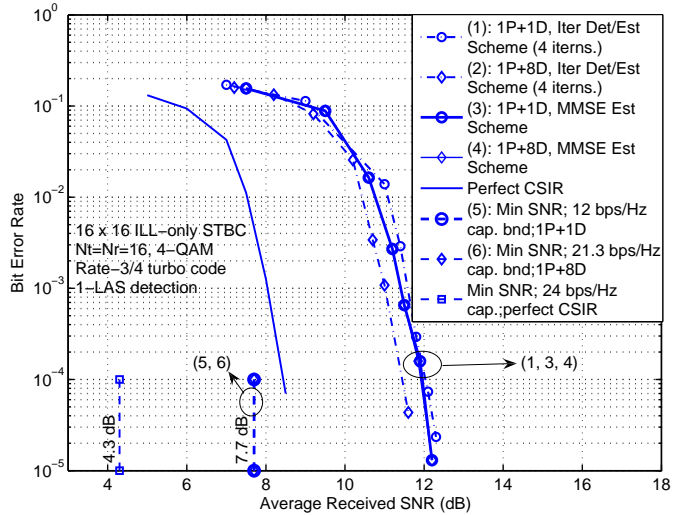}
\caption{
Turbo coded BER performance of 1-LAS detector for
$16 \times 16$ {\bf ILL-only} STBC with $i)$ perfect CSIR, $ii)$ CSIR
using MMSE estimation, and $iii)$ CSIR using iterative detection/channel
estimation (4 iterations). $N_t=N_r=16$, 4-QAM, rate-3/4 turbo code,
1P+1D \big($T=32, \tau=16, \beta_p=\beta_d=1$\big) and 1P+8D
\big($T=144, \tau=16, \beta_p=\beta_d=1$\big) training.
}
\vspace{-2mm}
\label{fig13}
\end{figure}

\begin{figure}
\hspace{-6mm}
\includegraphics[width=4.00in]{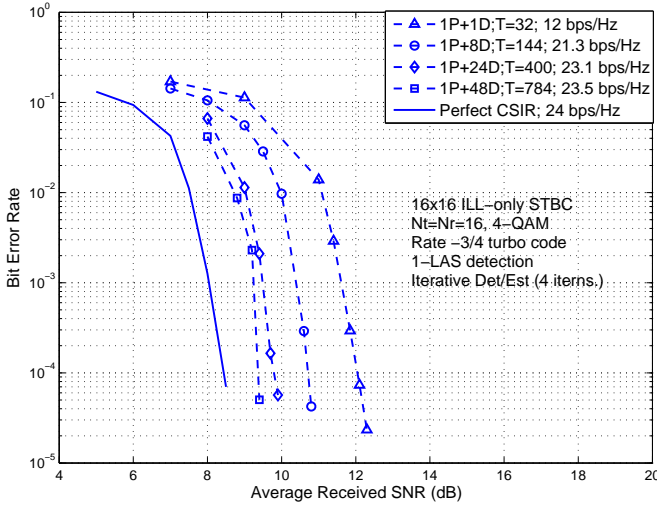}
\caption{
Turbo coded BER performance of 1-LAS detection and iterative
estimation/detection as a function of coherence time,
$T=32,144,400,784$, for a given $N_t=N_r=16$, $16\times 16$ {\bf ILL-only}
STBC, 4-QAM, rate-3/4 turbo code. {\em Spectral efficiency and BER
performance with estimated CSIR approaches to those with perfect CSIR
in slow fading (i.e., large $T$)}.
}
\vspace{2mm}
\label{fig14}
\end{figure}

\vspace{-2mm}
\hspace{-5mm}
where $T$ and $\tau$, respectively, are the frame size (i.e., channel
coherence time) and pilot duration in number of channel uses, and
$\sigma_{{\hat{\bf H}}_c}^2 = \frac{1}{N_tN_r} \ {\mathbb E} \big[ \mbox{tr}\{{\hat{\bf H}}_c {\hat {\bf H}}_c^H\} \big],$
where
$\hat{\bf H}_c = \ {\mathbb E} \big[{\bf H}_c\ \big|\ {\bf X}_c^{\text{(P)}},{\bf Y}_c^{\text{(P)}}\big]$
is the MMSE estimate of the channel gain matrix. We computed the capacity
bound in (\ref{trg_bound}) through simulations for 1P+8D and 1P+1D training
for a $16\times 16$ MIMO channel. For 1P+8D training $T=(1+8)16=144$,
$\tau=16$, and for 1P+1D training $T=(1+1)16=32$, $\tau=16$.  In
computing the bounds (shown in Fig. \ref{fig11}) and in BER simulations
(in Figs. \ref{fig12} and \ref{fig13}), we have used $\beta_p=\beta_d=1$.
In Fig. \ref{fig11},
we plot the computed capacity bounds, along with the capacity under perfect
CSIR \cite{tela99}. We obtain the minimum SNR for a given capacity bound in
(\ref{trg_bound}) from the plots in Fig. \ref{fig11}, and show (later in
Fig. \ref{fig11}) the nearness of the coded BER of the proposed scheme to
this SNR limit. We note that improved capacity and BER performance can be
achieved if optimum pilot/data power allocation derived in \cite{hh03}
is used instead of the allocation used in Figs. \ref{fig11} to
\ref{fig13} (i.e., $\beta_p=\beta_d=1$). We have used the optimum power
allocation in \cite{hh03} for
generating the BER plots in Figs. \ref{fig14} and \ref{fig15}.
In all the BER simulations with training, $\sqrt{\mu}\,{\bf I}_{N_t}$ is
used as the pilot matrix. ILL-only STBCs and 1-LAS detection are
used.

First, in Fig. \ref{fig12}, we plot the uncoded BER performance of 1-LAS
detector when 1P+1D and 1P+8D training are used for channel estimation
in a $16\times 16$ STBC MIMO system with $N_t=N_r=16$ and 4-QAM. BER
performance with perfect CSIR is also plotted for comparison. From
Fig. \ref{fig12}, it can be observed that, as expected, the BER degrades
with estimated CSIR compared to that with perfect CSIR. With MMSE estimation
scheme, the performance with 1P+1D and 1P+8D are same because of the one-shot
estimation. Also, with 1P+1D training, both the MMSE estimation scheme as
well as the iterative detection/estimation scheme (with 4 iterations between
detection and estimation) perform almost the same, which is about 3 dB worse
compared to that of perfect CSIR at an uncoded BER of $10^{-3}$. This
indicates that with 1P+$N_d$D training, iteration between detection and
estimation does not improve performance much over the non-iterative scheme
(i.e., the MMSE estimation scheme) for small $N_d$. With large $N_d$ (e.g.,
slow fading), however, the iterative scheme outperforms the non-iterative
scheme; e.g., with 1P+8D training, the performance of the iterative
detection/estimation improves by about 1 dB compared to the MMSE estimation.

Next, in Fig. \ref{fig13}, we present the rate-3/4 turbo coded BER of
1-LAS detector using estimated CSIR for the cases of 1P+8D and 1P+1D
training. From Fig. \ref{fig13}, it can be seen that, compared to that
of perfect CSIR, the estimated CSIR performance is worse by about 3 dB
in terms of coded BER for 1P+8D training. With MMSE estimation scheme,
$10^{-4}$ coded BER occurs at about $12-7.7=4.3$ dB away from the
capacity bound for 1P+1D and 1P+8D training. This nearness to capacity
bound improves by about 0.6 dB for the iterative detection/estimation
scheme. We note that for the system in Fig. \ref{fig13} with parameters
$16\times 16$ STBC, 4-QAM, rate-3/4 turbo code, and 1P+8D training with
$T=144,\tau=16,$ we achieve a high spectral efficiency of
$16\times 2\times \frac{3}{4} \times \frac{8}{9} = 21.3$ bps/Hz even
after accounting for the overheads involved in channel estimation (i.e.,
pilot matrix) and channel coding, while achieving good near-capacity
performance at low complexity. This points to the suitability of the
proposed approach of using LAS detection along with iterative
detection/estimation in practical implementation of large STBC MIMO
systems.

Finally, in Fig. \ref{fig14}, we illustrate the coded BER performance of
1-LAS detection and iterative detection/estimation scheme for different
coherence times, $T$, for a fixed $N_t=N_r=16$, $16\times 16$ STBC,
4-QAM, and rate-3/4 turbo code. The various values of $T$ considered and
the corresponding spectral efficiencies are: $i)$ $T=32$, 1P+1D, 12 bps/Hz,
$ii)$ $T=144$, 1P+8D, 21.3 bps/Hz, $iii)$ $T=400$, 1P+24D, 23.1 bps/Hz, and
$iv)$ $T=784$, 1P+48D, 23.5 bps/Hz. In all these cases, the corresponding
optimum pilot/data power allocations in \cite{hh03} are used. From Fig.
\ref{fig14}, it can be
seen that for these four cases, $10^{-4}$ coded BER occurs at around 12 dB,
10.6 dB, 9.7 dB, and 9.4 dB, respectively. The $10^{-4}$ coded BER for
perfect CSIR happens at around 8.5 dB. This indicates that the performance
with estimated CSIR improves as $T$ is increased, and that a performance loss
of less than 1 dB compared to perfect CSIR can be achieved with large $T$
(i.e., slow fading). For example, with 1P+48D training ($T=784$), the
performance with estimated CSIR gets close to that with perfect CSIR both
in terms of spectral efficiency (23.5 $vs$ 24 bps/Hz) as well as SNR at
which $10^{-4}$ coded BER occurs (8.5 $vs$ 9.4 dB). This is expected,
since the channel estimation becomes increasingly accurate in slow fading
(large coherent times) while incurring only a small loss in
spectral efficiency due to pilot matrix overhead. This result is
significant because $T$ is typically large in fixed/low-mobility wireless
applications, and the proposed system can effectively achieve high spectral
efficiencies as well as good performance in such applications.

\begin{table}
\begin{center}
\begin{tabular}{|l|c|c|}
\hline
Parameters                  & System-I           & System-II         \\ \hline
\hline
\# Rx antennas, $N_r$       & 16                 & 16                \\
Coherence time, $T$         & 48                 & 48                \\
{\bf \# Tx antennas, $N_t$} & {\bf 16}           & {\bf 12}          \\
STBC from CDA               & $16\times 16$      & $12\times 12$     \\
Pilot duration, $\tau$      & 16                 & 12                \\
Training                    & 1P+2D              & 1P+3D             \\
$\beta_p^{opt}$             & 1.2426             & 1.4641            \\
$\beta_d^{opt}$             & 0.8786             & 0.8453            \\
Modulation                  & 4-QAM              & 4-QAM             \\
Turbo code rate             & $1/2$              & $3/4 $            \\
{\bf Spectral efficiency}   & {\bf 10.33} bps/Hz & {\bf 13.5} bps/Hz \\
{\bf SNR at $10^{-3}$ coded BER} & {\bf 8.9} dB  & {\bf 8.6} dB     \\
\hline
\end{tabular}
\label{tab3}
\caption{
On optimum $N_t$ for a given $N_r$ and $T$. System-II with a
smaller $N_t$ achieves a higher spectral efficiency while achieving
$10^{-3}$ coded BER at a lesser SNR than System-I with a larger $N_t$. }
\end{center}
\vspace{-4mm}
\end{table}

\subsection{On Optimum $N_t$ for a Given $N_r$ and $T$}
\label{secc}
\begin{figure}
\hspace{-6mm}
\includegraphics[width=4.00in]{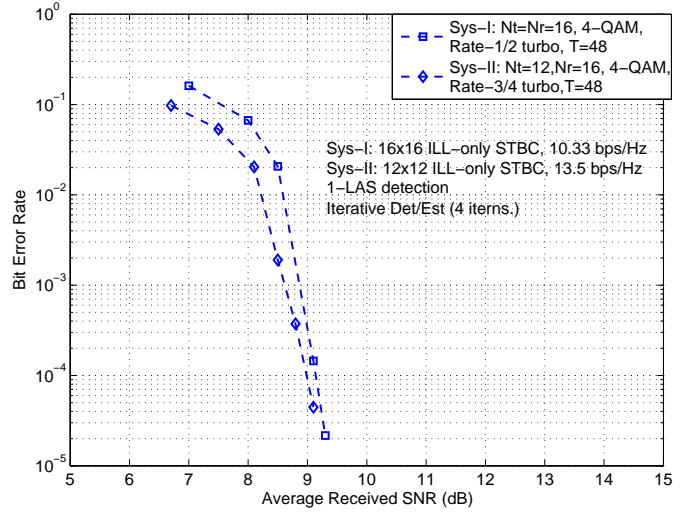}
\caption{
Comparison between two 1P+$N_d$D training-based systems, one with a
larger $N_t$ than the other for a given $N_r$ and $T$. {\em With $N_r=16$,
$T=48$ and optimum power allocation in both systems,
System-II with $N_t=12$ achieves a higher spectral efficiency
\big(13.5 vs 10.33 bps/Hz\big) while achieving $10^{-3}$ coded BER at a
lesser SNR \big(8.6 vs 8.9 dB\big) than System-I with $N_t=16$.}
}
\vspace{-2mm}
\label{fig15}
\end{figure}

In \cite{hh03}, through theoretical capacity bounds it has been shown
that, for a given $N_r$, $T$ and SNR, there is an optimum value of $N_t$
that maximizes the capacity bound \big(refer Figs. 5 and 6 in \cite{hh03},
where the optimum $N_t$ is shown to be greater than $N_r$ in Fig. 5 and
less than $N_r$ in Fig. 6\big). For example, for $N_r=16$, $T=48$, and
SNR = 10 dB, the capacity bound evaluated using (\ref{trg_bound}) with
optimum power allocation for $N_t=12$ is 19.73 bps/Hz, whereas for
$N_t=16$ the capacity bound reduces to 17.53 bps/Hz showing that the
optimum $N_t$ in this case will be less than $N_r$. We demonstrate such
an observation in practical systems by comparing the simulated coded BER
performance of two systems, referred to as System-I and System-II,
using 1-LAS detection and iterative detection/estimation scheme.
The parameters of System-I and System-II are listed in Table II.
$N_r$ and $T$ are fixed at 16 and 48, respectively, in both systems.
System-I uses 16 transmit antennas and $16\times 16$ STBC,
whereas System-II uses 12 transmit antennas and $12\times 12$
STBC. Since the pilot matrix is $\sqrt{\mu}\,{\bf I}_{N_t}$, the pilot
duration $\tau$ is 16 and 12, respectively, for System-I and System-II.
Optimum pilot/data power allocation and 4-QAM modulation are employed in
both systems. System-I uses rate-1/2 turbo code and system-II uses rate-3/4
turbo code. With the above system parameters, the spectral efficiency
achieved in System-I is
$16\times 2\times \frac{1}{2}\times \frac{2}{3} = 10.33$ bps/Hz,
whereas System-II achieves a higher spectral efficiency of
$12\times 2\times \frac{3}{4}\times \frac{3}{4}= 13.5$ bps/Hz.
In Fig. \ref{fig15}, we plot the coded BER of both these
systems using 1-LAS detection and iterative detection/estimation.
From the simulation points shown in Fig. \ref{fig15}, it can be
observed that System-II with a smaller $N_t$ and higher spectral
efficiency in fact achieves a certain coded BER performance at a
lesser SNR compared to System-I. For example, to achieve $10^{-3}$
coded BER, System-I requires an SNR of about 8.9 dB, whereas System-II
requires only 8.6 dB. This implies that because of the reduction of
throughput due to pilot symbols \big(by a factor of $\frac{T-\tau}{T}$
for a given $T$ and $\tau=N_t$\big), a larger $N_t$ does not necessarily
mean a higher spectral efficiency. Such an observation has also been
made in \cite{hh03} based on theoretical capacity bounds. The proposed
detection/channel estimation scheme allows the prediction of such
behavior through simulations, which, in turn, allows system designers
to find optimum $N_t$ and STBC size to achieve a certain spectral
efficiency in large STBC MIMO systems.

\section{Conclusion}
\label{sec6}
We presented a low-complexity algorithm for the detection of high-rate,
non-orthogonal STBC large-MIMO systems with tens of antennas that achieve
high spectral efficiencies of the order of several tens of bps/Hz. We also
presented a training-based iterative detection/channel estimation scheme
for such large STBC MIMO systems. Our simulation results showed that the
proposed 1-LAS detector along with the proposed iterative detection/channel
estimation scheme achieved very good performance at low complexities. With
the feasibility of low-complexity high-performance receivers, like the
proposed detection/channel estimation scheme, large-MIMO systems with tens
of antennas at high spectral efficiencies can become practical, enabling
interesting high data rate wireless applications (e.g., wireless IPTV/HDTV
distribution). This can motivate the inclusion of large-MIMO architectures
(e.g., $12\times 12$, $16\times 16$ MIMO systems, including those using
STBCs from CDA) into wireless standards like IEEE 802.11n/VHT and IEEE
802.16/LTE-A in their evolution to achieve high data rates at
increased spectral efficiencies.

\section*{Appendix}
\begin{thm}
The $l_p^{(k)}$ in (\ref{Optdp}) minimizes {\small
$\mathcal F(l_p^{(k)})$} in (\ref{CostDiff2}) and this minimum value is
non-positive.
\end{thm}
{\em Proof:}
Let
$r \Define \left\lfloor \frac{\vert z_p^{(k)} \vert}{2 a_p} \right\rfloor$.
Then
$\frac{\vert z_p^{(k)} \vert}{2 a_p} = r + f$,  where $0 \leq f < 1$,
and so we can write
\begin{eqnarray}\label{ProofOpt2}
\frac { \vert z_p^{(k)} \vert } { a_p} & = & 2r + 2f .
\end{eqnarray}
If $l_p^{(k)}$ were unconstrained to be any real number, then the optimal
value of $l_p^{(k)}$ is $\frac {\vert z_p^{(k)} \vert }{ a_p}$, which would
lie between $2r$ and $2r+2$ (as per (\ref{ProofOpt2})). Since
{\small $\mathcal F(l_p^{(k)})$} is quadratic in $l_p^{(k)}$, it is
unimodular, and hence the optimal point (with $l_p^{(k)}$ constrained)
would be either $2r$ or $2r+2$. Using
(\ref{CostDiff2}) and (\ref{ProofOpt2}), we can evaluate
$\mathcal F(2r+2) - \mathcal F(2r)$ to be
\begin{eqnarray}
\label{ProofOpt3}
\mathcal F(2r+2) - \mathcal F(2r)
& = & 4a_p( 1 - 2f).
\end{eqnarray}
Since $a_p$ is a positive quantity, the sign of
$\mathcal F(2r+2)- \mathcal F(2r)$ depends upon the sign of
$(1-2f)$. If $f \geq 0.5$, then $\mathcal F(2r+2) \leq \mathcal F(2r)$, and
therefore $2r+2$ is the optimal value of $l_p^{(k)}$. Similarly, when
$f < 0.5$, $2r$ is the optimal value of $l_p^{(k)}$. Therefore, it
follows that indeed the rounding solution given by (\ref{Optdp}) is optimal.
{\small $\mathcal F(l_p^{(k)})$} is non-positive
for all values of $l_p^{(k)}$ between zero and
$\frac{2 \vert z_p^{(k)} \vert}{a_p}$. If $f < 0.5$, then $2r$ is
optimal, and, from (\ref{ProofOpt2}), we know that
$2r \leq \frac { \vert z_p^{(k)} \vert }{a_p}$, and therefore
$2r < 2\frac { \vert z_p^{(k)} \vert }{a_p}$.
Hence $\mathcal F\left(2r\right)$ $= \mathcal F^{(opt)}$ is non-positive.
Similarly, if $f \geq 0.5$, then $2r+2$ is optimal, and
$\mathcal F(2r+2) \leq \mathcal F(2r)$. However, since $2r$ is
always less than $2\frac{\vert z_p^{(k)} \vert}{a_p}$,
$\mathcal F(2r)$ is non-positive and therefore
$\mathcal F(2r+2)$ $= \mathcal F^{(opt)}$ is non-positive.


\section*{ACKNOWLEDGMENT}
We would like to thank the Editor, Prof. R. Calderbank, for handling 
the review process. 
We would like to thank the anonymous reviewers for their critical and
useful comments, and for motivating us to compare the performance and 
complexity of the proposed scheme with those of other large-MIMO 
architectures/detectors.

\vspace{-4mm}
\begin{IEEEbiography}[]{Saif K. Mohammed}
received his B.Tech degree in Computer Science and
Engineering from the Indian Institute of Technology, New Delhi, India,
in 1998. From 1998 to 2000, he was employed with Philips Inc., Bangalore,
as an ASIC design engineer. From 2000 to 2003, he worked with Ishoni
Networks Inc., Santa Clara, CA, as a senior chip architecture engineer.
From 2003 to 2007, he was employed with Texas Instruments, Bangalore as
systems and algorithms designer in the wireless systems group. Since 2006,
he is pursuing his doctoral degree in Electrical and Communications
Engineering at the Indian Institute of Science, Bangalore, India. His
research interests include low-complexity detection, estimation and
coding for wireless communications systems.
\end{IEEEbiography}

\begin{IEEEbiography}[]{Ahmed Zaki}
received the B.E. degree in Electronics and Communication Engineering 
from Osmania University, Hyderabad, India, in 2007, and the M.E. degree 
in Telecommunication from the Indian Institute of Science, Bangalore, 
India, in 2009. His research interest lies in the area of wireless 
communications, including receiver design and channel estimation for 
large-MIMO systems, MIMO-OFDM, multiuser communications, and algorithm 
design.
\end{IEEEbiography}

\begin{IEEEbiography}[]{A. Chockalingam}
was born in Rajapalayam, Tamil Nadu, India. He
received the B.E. (Honors) degree in Electronics and Communication
Engineering from the P. S. G. College of Technology, Coimbatore,
India, in 1984, the M.Tech degree with specialization in satellite
communications from the Indian Institute of Technology, Kharagpur,
India, in 1985, and the Ph.D. degree in Electrical Communication
Engineering (ECE) from the Indian Institute of Science (IISc),
Bangalore, India, in 1993. During 1986 to 1993, he worked with the
Transmission R \& D division of the Indian Telephone Industries
Limited, Bangalore. From December 1993 to May 1996, he was a
Postdoctoral Fellow and an Assistant Project Scientist at the
Department of Electrical and Computer Engineering, University of
California, San Diego. From May 1996 to December 1998, he served
Qualcomm, Inc., San Diego, CA, as a Staff Engineer/Manager in the
systems engineering group. In December 1998, he joined the faculty
of the Department of ECE, IISc, Bangalore, India, where he is a
Professor, working in the area of wireless communications and 
networking.

Dr. Chockalingam is a recipient of the Swarnajayanti Fellowship from
the Department of Science and Technology, Government of India. He
served as an Associate Editor of the IEEE Transactions on Vehicular
Technology from May 2003 to April 2007. He currently serves as an
Editor of the IEEE Transactions on Wireless Communications. 
He served as a Guest Editor for the IEEE JSAC Special Issue on
Multiuser Detection for Advanced Communication Systems and Networks.
He is a Fellow of the Institution of Electronics and Telecommunication 
Engineers, and a Fellow of the Indian National Academy of Engineering.
\end{IEEEbiography}

\begin{IEEEbiography}[]{B. Sundar Rajan}
(S'84-M'91-SM'98) was born in Tamil Nadu, India. He received the B.Sc. degree 
in mathematics from Madras University, Madras, India, the B.Tech degree in 
electronics from Madras Institute of Technology, Madras, and the M.Tech and 
Ph.D. degrees in electrical engineering from the Indian Institute of 
Technology, Kanpur, India, in 1979, 1982, 1984, and 1989 respectively. He 
was a faculty member with the Department of Electrical Engineering at the 
Indian Institute of Technology in Delhi, India, from 1990 to 1997. Since 
1998, he has been a Professor in the Department of Electrical Communication 
Engineering at the Indian Institute of Science, Bangalore, India. His 
primary research interests include space-time coding for MIMO channels, 
distributed space-time coding and cooperative communication, coding for 
multiple-access, relay channels and network coding with emphasis on 
algebraic techniques.

Dr. Rajan is an Associate Editor of the IEEE Transactions on Information 
Theory, an Editor of the IEEE Transactions on Wireless Communications, and 
an Editorial Board Member of International Journal of Information and Coding 
Theory. He served as Technical Program Co-Chair of the IEEE Information 
Theory Workshop (ITW'02), held in Bangalore, in 2002. He is a Fellow of 
Indian National Academy of Engineering and recipient of the IETE Pune 
Center's S.V.C Aiya Award for Telecom Education in 2004. Also, Dr. Rajan 
is a Member of the American Mathematical Society.
\end{IEEEbiography}

\vfill
\end{document}